\def\eq#1{equation~(#1)}

\def\fig#1{Figure~\ref{#1}}

\documentstyle[emulateapj]{article}

\def\gsim{{{}_>\atop{}^{{}^\sim}}}
\def\lsim{{{}_<\atop{}^{{}^\sim}}}

\def\dol{{D_{\rm OL}}}
\def\dos{{D_{\rm OS}}}
\def\dls{{D_{\rm LS}}}
\def\umin{{u_{\rm min}}}
\def\te{t_E}
\def\kpc{~{\rm kpc}}
\def\kms{~{\rm km\, s^{-1}}}
\def\days{~{\rm days}}
\def\dchi{\Delta\chi^2}
\def\dchit{\Delta\chi^2_{\rm thresh}}
\def\elz{{\bar \epsilon}_{\rm LZ}}

\begin{document}

\submitted{ Submitted to ApJ April 23, 1999}

\title{Detection Efficiencies of Microlensing Datasets to Stellar and  
Planetary Companions}

\author{B. Scott Gaudi}

\affil{Ohio State University, Department of Astronomy, Columbus, OH 43210 \\
gaudi@astronomy.ohio-state.edu}

\author{Penny D. Sackett}

\affil{Kapteyn Astronomical Institute, 9700 AV Groningen, The
  Netherlands \\ psackett@astro.rug.nl}

\begin{abstract}

Microlensing light curves are now being monitored 
with the temporal sampling and photometric precision required 
to detect small perturbations due to planetary companions of the primary 
lens.  Microlensing is complementary to other planetary  
search techniques, both in the mass and orbital separation of the planets to 
which it is sensitive and its potential for measuring the statistical 
frequency of planets beyond the solar neighborhood.   
We present an algorithm to analyze the efficiency with which  
the presence of lensing binaries of given mass ratio 
and angular separation can be detected in real microlensing datasets.   
Such an analysis is required in order to draw 
statistical inferences about lensing companions, 
and differs from previous studies of idealized microlensing 
experiments by incorporating  
instead the actual sampling, photometric precision, and monitored 
duration of individual light curves. 
We apply the method to artificial (but realistic) data to explore 
the dependence of detection efficiencies on observational parameters, 
the impact parameter of the event, the finite size of the background source, 
the amount of unlensed (blended) light, and the criterion used to 
define a detection.  We find that: 
(1) the integrated efficiency depends strongly on the impact 
parameter distribution of the monitored events, 
(2) calculated detection efficiencies are robust to changes in 
detection criterion for strict criteria ($\Delta\chi^2 \gsim 100$) and 
large mass ratios $(q\gsim 10^{-2}$), 
(3) finite sources can dramatically alter detection efficiencies to 
companions with mass ratios $q \lsim 10^{-3}$, and 
(4) accurate determination of the blended light fraction 
is crucial for the accurate determination of the 
detection efficiency of individual events.  
Suggestions are given for addressing complications associated with computing 
accurate detection efficiencies of real datasets.

\end{abstract}

 
\section{Introduction}

The discovery in 1995 of a massive planet orbiting 51 Peg 
(\cite{mandq1995}) followed by the discovery of several more planets 
orbiting nearby dwarf stars using the same radial velocity technique 
(see \cite{mandb1998} and references therein) has focussed both 
public and scientific attention on the search for extra-solar planets 
and the experimental and theoretical progress being made in developing 
other viable detection techniques.  

Due to their small mass and size, extra-solar planets are difficult to
find.  Proposed detection methods can be
subdivided into direct and indirect techniques.  Direct
methods rely on the detection of the reflected light of the parent star, 
and are exceedingly challenging due to the extremely small flux
expected from the planet, which is overwhelmed by stray light from the
star itself (\cite{aandw1997}).  
Some direct imaging searches have already been performed
(\cite{boden1998b}), but the future of this method lies in the
construction and launching of a space-based interferometer (\cite{wanda1998}).

Astrometric, radial velocity and occultation measurements can be used 
to detect the presence of a planet indirectly.
Astrometric detection relies on the measurement of the wobble of the
stellar centroid caused by the motion of the star around the
center-of-mass of the planet-star system, and yields the 
mass ratio of the planet-star system.  Many attempts to find extra-solar
planets in this way have been made, but the measurements are 
difficult and the detections remain controversial; 
planned space-based missions are expected 
to be more successful (\cite{linde96}).  Occultation methods use very accurate
photometry of the parent star to detect the small decrease in flux
($\lsim 1\%$) caused by a planet transiting 
the face of the star (\cite{handd1994}).
Occultation searches are currently being carried out (\cite{deeg}), 
and spaced-based missions are planned to increase the sensitivity  
to small-mass planets (\cite{corot}, \cite{kepler}).  
By far the most successful indirect method for discovering planets has been
the Doppler technique, which employs extraordinarily precise
radial velocity measurements of nearby stars
to detect Doppler shifts caused by orbiting planets.  
Several teams have monitored nearby stars with the aim of detecting the 
Doppler signal of orbiting planets
(\cite{mcmilan1993}; \cite{mandq1995}; \cite{butler1996}, \cite{cochran1997}, \cite{noyes1997}). To date these groups combined have discovered over 
20 extra-solar planets using this technique, with new planetary companions 
being announced every few months (\cite{mandb1998}).  
Of these techniques, 
only the proposed space-based interferometric imaging and transit searches 
are expected to be sensitive to Earth-mass planets.

These detection techniques are complementary to one another both 
in terms of their sensitivity to planetary mass and orbital separations 
and the specific physical quantities of the planetary system that they measure.  
All share two distinct advantages: 
the experiments are repeatable and, due to their reliance on 
flux measurements of the parent star or the planet itself, they are 
sensitive to stars in the solar neighborhood where 
follow-up studies can be most easily pursued.  For example, 
spectroscopic follow-up studies may enable the detection of molecules 
commonly thought to be indicative of life, such as water, carbon dioxide, 
and ozone (\cite{wanda1998}).
This advantage is linked to a common drawback: 
the searches can be conducted only on a limited number of nearby stars.  
Without substantial developments in technology beyond what
is currently available, these methods are unlikely to detect 
planets orbiting stars more distant than $\sim$1$\,{\rm kpc}$, 
and are thus unable to address questions about the nature of 
planetary systems beyond the immediate solar neighborhood.

Microlensing was proposed in 1986 by Paczy{\'n}ski as a method to detect
compact baryonic dark matter in the halo of our Galaxy.  Soon after,
three collaborations (MACHO, \cite{alcock1993}; EROS, \cite{aubourg1993}; 
OGLE, \cite{udalski1993}) were organized to implement this
proposal with massive observational programs.  
The basic idea is simple: when a compact object, such as a star or 
massive dark object, passes near the line of sight to a
distant source, the gravitational field of the intervening object will
serve as a lens, creating two images of the distant source.  For typical
stellar sources and lenses in our Galaxy, the separation of the images 
will be on the order of milliarcseconds and hence unresolved. 
However, the lens will also magnify the source; this magnification is
measurable and depends (only) on the angular separation 
of the lens and source.  Since the lens, source, and
observer are all in relative motion, the magnification will be
time-variable, creating a ``microlensing event.''  Because the probability
that any single source star will be microlensed (the `optical depth') is
so low, ${\cal O} (10^{-6})$, millions of stars must be monitored;
the crowded fields towards the Large and Small Magellenic clouds 
(LMC and SMC) and the Galactic bulge were thus the natural targets.  
In 1993, the first candidate microlensing events were announced toward the
LMC (\cite{alcock1993}; \cite{aubourg1993}) and Galactic bulge
(\cite{udalski1993}).  Six years later, these three 
`survey teams' continue to search and 
discover microlensing events. Over $400$ events have been discovered
toward all three targets (MACHO, Alcock et~al.\ 1997a, 1997b, 1997c;
EROS, \cite{renault1998}, \cite{pdeb1998}; OGLE, \cite{udalski1997}), 
the overwhelming majority of which are discovered toward the Galactic bulge.  

Although Paczy{\'n}ski's original suggestion was to search for dark
matter, before the first event was discovered Mao \& Paczy{\'n}ski (1991) 
had already noted that it might be
possible to detect planetary companions of the primary microlenses via the 
distortions they create in the magnification pattern, 
and thus the light curve, generated by the primary lens.  
The nature of the distortion depends on the mass 
ratio and angular separation of the two components. 

As a planet search technique, microlensing offers unique advantages.  
Since microlensing is caused the gravitational field (i.e., mass) 
of the lenses, it is not limited to the study of 
nearby or indeed luminous objects, and thus can be used to search 
for planetary companions around typical Galactic stars at distances 
of many kiloparsecs.  
As a consequence, a nearly unlimited number of dwarf stars are available 
to serve as gravitational microlenses and potential search candidates.
This advantage is linked to the primary drawback of microlensing 
planetary searches: most follow-up studies will be difficult due to the 
faintness of the stars serving as typical lenses.  
This drawback is compounded by the irrepeatability of 
specific microlensing observations; lensing
of a particular source by a particular lens is a singular occurrence.
Nevertheless, the robust statistics on the nature of 
planetary systems many kiloparsecs distant and the complementary 
nature of the information about discovered systems 
microlensing can provide make it an important tool in the 
cadre being assembled to study extra-solar planets (\cite{sackett1999a}). 
Furthermore, microlensing planet searches are relatively
inexpensive, requiring only several dedicated $1$-meter class
telescopes.  Microlensing is the only technique currently capable 
of routinely discovering planets like our own Jupiter, 
and the only ground-based method capable, in principle, 
of detecting distant terrestrial-mass planets, 
though this will require substantial enhancements over existing 
capabilities (\cite{bandr1996}; \cite{peale1997}, \cite{sackett1997}).

Current microlensing survey teams have sampling rates that are
too large ($\sim$day) and/or photometric accuracies that are too poor
($\sim 5\%$) to detect and characterize these perturbations, but  
the nearly 100 real-time electronic alerts of on-going bulge events 
that they provide annually have become the 
primary targets for newly-formed microlensing planet searches. 
These new `monitoring teams' 
(PLANET, Albrow et~al.\ 1997, 1998; GMAN, \cite{alcock1997d}; 
MPS, \cite{mps}) have formed with the express purpose of 
executing the nearly continuous temporal coverage and
high photometric precision on real-time microlensing alerts 
necessary to detect deviations from
the generic light curve of the sort expected from planetary
and other microlensing anomalies (see e.g.\ \cite{albrow1998}).
The PLANET collaboration in particular has now monitored 
nearly 100 microlensing events with varying degrees of
photometric sampling and precision (often $\sim$hourly with 
$\sim$2\% over the largest magnification regions); such
existing datasets may already place interesting constraints on the
frequency of stellar binaries and planetary systems.

Like all planet search techniques, microlensing is not $100\%$
efficient, due to both intrinsic and observational limitations.  
The efficiency with which a given dataset will reveal the presence of
a companion to the primary microlens must be quantified before it can be used to 
constrain the frequency and properties of extra-solar planetary systems.
Quantification of detection efficiencies of any
kind can be a difficult and tedious process: 
the intrinsic limitations of the method must be identified and 
combined with the actual observational limitations.  
The detection efficiency may depend on hidden or unmeasurable parameters; 
these must be identified and properly quantified in order to avoid 
biasing the final conclusions. Detection efficiencies of 
other planet search techniques have been presented and
applied to data (\cite{nanda1998}).  However, despite a substantial body 
of work addressing the likely planet detection 
efficiency of idealized microlensing programs (see \S\ 3),  
no methods have yet been proposed for calculating the 
efficiency of actual datasets to lensing binaries of a given type.

Here we present an algorithm for computing the 
detection efficiency to lensing binaries that is specific to 
individual microlensing light curves.  
By directly imposing the actual observational limitations as a constraint, 
the approach is less prone to the biases that may arise when using 
simplified models of observational conditions.  
Furthermore, the method is simple to implement and computationally 
inexpensive, since it involves direct integration over 
unknown quantities rather than Monte Carlo simulations commonly used 
to calculate detection efficiencies of idealized observing programs. 
We apply this method to simulated datasets in order to explore how 
the detection efficiency depends on intrinsic and observational
effects.  We also explore possible biases that may be introduced into the
inferred efficiency of individual events to planet detection 
if the size of the source and the fraction of unresolved light (`blending') 
are ill-constrained. 

In \S\ 2, we review the relevant formalism for
microlensing by single and double stars.  
A brief review of the literature on expectations for idealized microlensing 
planet searches is presented in \S\ 3, contrasting these to the goal of 
this paper.  Detection and detection efficiency, as used throughout 
this work, are defined in \S\ 4, along with a general overview of the 
connection between detection efficiency 
and the mass ratio, angular separation and impact parameter of the 
event.  Our algorithm for computing detection 
efficiencies is described in \S\ 5 and applied to artificial data 
in order to access the effects of different detection criterion.  
The effects of finite source size on detection efficiencies are  
presented in \S\ 6; the effects of blending in \S\ 7.  
Suggestions for addressing complications associated with computing 
accurate detection efficiencies of real microlensing datasets 
are given in \S\ 8.  
We summarize and conclude in \S\ 9.

\begin{figure*}[t]
\epsscale{1.2}
\plotone{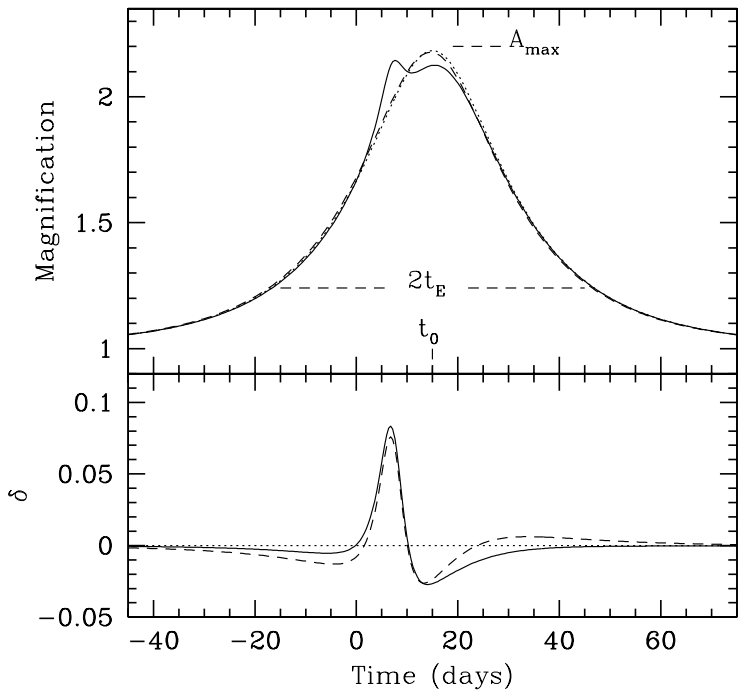}
\caption{
\footnotesize
Top panel: Magnification as a
function of time in days for a microlensing event with (solid curve)
and without (dotted curve) the presence of a companion with mass ratio $q=10^{-2}$.
The event has an Einstein time scale of $\te=30~{\rm{days}}$ and minimum
impact parameter in units of the Einstein ring radius of $\umin=0.5$.  
The single lens curve reaches a maximum magnification
of $A_{\rm max}=2.2$ at a time $t_0=15~{\rm{days}}$.  The binary lens
curve was generated assuming that the source trajectory makes angle 
$\theta=300^{\circ}$ with respect to the binary axis.  
Also shown is the best single-lens fit to the binary-lens
light curve (dashed curve). 
Bottom panel: The fractional
deviation $\delta$ of the binary-lens
light curve from the single-lens light curve with parameters
$\te=30~{\rm{days}}, \umin=0.5, t_0=15~{\rm days}$ (solid curve) and
from the best fit single-lens light curve (dashed curve). }
\label{fig:defs}
\end{figure*}

\section{Relevant Formalism for Single and Double Microlenses}

\subsection{Single lenses}
The time-variable flux observed from a microlensed star is 
\begin{equation}
F(t)=F_0 [A(t)+f_B]~~~,
\label{fluxt}
\end{equation}
\noindent
where $F_0$ is the unlensed flux of the star, $f_B$ is the ratio of any
unresolved, unlensed background light to $F_0$ (the `blend
fraction'), and $A(t)$ is the magnification.  The magnification of 
$A_0$ of a point source by a point lens can be written as  
\begin{equation}
A_0 = {{u^2+2}\over u (u^2+4)^{1/2}}~~.
\label{mag0}
\end{equation}
Here $u$ is the instantaneous angular separation of the source and the 
lens in units of the angular Einstein ring radius $\theta_E$ of the lens, 
a degenerate combination of the lens mass and and distances defined by 
\begin{equation}
\theta_E\equiv\left[{4GM \over c^2} {D_{LS} \over D_{OL} \, D_{OS}}\right]^{1/2}
\sim 130 \, \mu{\rm as} \left({M \over 0.2 M_{\odot}}\right)^{1/2},
\label{thetae}
\end{equation}
where $M$ is the mass of the lens, and $\dls$, $\dos$, $\dol$ are the
lens-source, observer-source, and observer-lens distances, respectively.
For the scaling relation on the far right of \eq{\ref{thetae}}, we have
assumed $\dol=6\kpc$ and $\dos=8\kpc$.  Note that for $u\rightarrow 0$,
$A_0 \rightarrow 1 /u$.  Since the source, lens, and observer are all in
relative motion, $u$ will be a function of time.  For rectilinear
motion, it can be shown that
\begin{equation}
u(t)= \left[ \left({t-t_0 \over \te}\right)^2 + u^2_{\rm
min}\right]^{1/2},
\label{uoft}
\end{equation}
where $t_0$ is the time of maximum magnification, $\umin$ is the minimum
angular separation, or impact parameter, of the event 
in units of $\theta_E$, and $\te$ is a characteristic 
time scale of the event, the Einstein time, defined by 
\begin{equation}
\te= { \theta_E \dol \over v_{\perp} } \sim 10\days \left({M\over 0.2 M_{\odot}}\right)^{1/2}.
\label{timescale}
\end{equation}
Here $v_{\perp}$ is the transverse velocity of the lens relative to the
observer-source line-of-sight.  For the scaling relation on the far 
right of \eq{\ref{timescale}}, we have assumed $\dol=6\kpc$, $\dos=8\kpc$, and
$v_{\perp} = 100 \kms$.

A point-lens point-source (PSPL) light curve is thus a function of five
parameters, $t_0$, $\te$, $\umin$, $F_0$, and $f_B$.  
Unless the lens is measurably luminous, 
the parameters $F_0$ and $f_B$ depend only on the source and its
environment.  The distributions of $F_0$ and $f_B$ depend on 
the luminosity functions of the observed sources and any blended light. 
The parameters $t_0$ and $\umin$ are purely geometrical.
Since the distribution of $t_0$ is flat and its value has no effect on
the analysis of detection efficiencies, we will hereafter set $t_0=0$.
The intrinsic distribution of $\umin$ is also flat, but the observed 
distribution has an upper limit set by the detection threshold 
of the survey teams.  If the lens is not contributing significantly 
to the blended light, 
only the characteristic time $\te$ contains physical information 
about the lens itself.  Its intrinsic distribution is set by 
the lens masses and spatial distribution of the lenses and 
sources; the observed distribution of $\te$ 
depends on the temporal sampling of the microlensing campaigns.  
An example of a point-lens light curve with $t_0=15\days$,
$\te=30 \days$, $\umin=0.5$, $F_0=1$, and $f_B=0$ is shown as the dotted
line in the top panel of \fig{fig:defs}.

\subsection{Binary lenses}

A companion orbiting a lens constitutes a double lens and 
thus can be described by the formalism of binary lenses.  
The flux is still expressed by \eq{\ref{fluxt}},
but the magnification can no longer be calculated analytically.  
Instead, the lens equation describing 
the mapping from the source plane ($\xi,\eta$) to the image plane
($x,y$) must be solved.  Following Witt (1990), we 
write the lens equation for two masses with 
fractional mass $m_1$ and $m_2$ located at positions 
$z_1$ and $z_2$ in terms of the complex coordinates 
$\zeta \equiv \eta+i\xi$ and $z\equiv x+iy$, 
\begin{equation}
\zeta = z + { m_1 \over \bar{z}_1 - \bar{z} } + {m_2 \over \bar{z}_2 - \bar{z} }~~.
\label{lenseq}
\end{equation}
Here all distances are in units of $\theta_E$, the Einstein ring radius of the
total mass $M$ of the binary.  The mapping is described completely 
by two parameters, the instantaneous angular separation of the two components 
in units of $\theta_E$, $b=|z_1-z_2|$, and the mass ratio of the system,
$q=m_2/m_1$.  Without loss of generality, we will assume that $m_2 \le
m_1$, so that $q \le 1$.  
Equation (\ref{lenseq}) is equivalent to a fifth-order
complex polynomial in $z$, which can be solved by the usual techniques.  
Each source position produces either three or five images.     
The magnification $A_j$ of each image $j$ is inversely proportional to the
determinant of the Jacobian of the lens mapping, evaluated at that
image position,
\begin{equation}
A_j = \left.{1\over{ |{\rm{det}} J|}}\right|_{z=z_j},\, \, \,
{\rm{det}}J = 1 - {{\partial\zeta}\over{\partial{\bar z}}}
\overline{{{\partial\zeta}\over{\partial{\bar z}}}}.
\label{magimage}
\end{equation}
The total magnification is given by the sum of the individual
magnifications, $A = \sum_j A_j$. The set of source positions for which
the magnification is formally infinite, given by the condition
${\rm{det}} J = 0$, defines a set of closed curves called caustics. 
Depending on the values of $b$ and $q$, a binary lens may have one, two, 
or three caustics; the multiplicity of images changes by two 
as the source crosses a caustic.

A static, point-source binary-lens light curve is a function of eight
parameters.  Two are identical to the point-lens case, $F_0$ and $f_B$.
The equation for $\te$ retains the same form, 
but the choice of the fiducial mass $M$ is arbitrary and 
can refer to the total mass of the binary or the mass of
one of the components.  For binaries, $\umin$ refers to the
minimum angular separation (in units of $\theta_E$) 
between the source and the origin of the binary system.  
The choice of origin is also arbitrary; popular choices
are the position of center of mass, the position of one of the masses,
or the midpoint between the two.  Equation (\ref{uoft}) still holds for static
binaries, so that $t_0$ is the time at which $u=\umin$, but 
for binary lenses this need not be the time of maximum magnification.  
The mapping parameters $b$ and $q$, and the angle $\theta$ on the sky 
between source trajectory and the binary axis, are the final 
three binary lens parameters.  In the case of a single lens, the
lensing geometry is azimuthally symmetric, and $\theta$ is completely
degenerate for any measured light curve.

The value of $\umin$ has a large effect on the detection efficiency of a 
given light curve to lens binarity; smaller $\umin$ events generally 
have higher efficiency.  
The Einstein time scale $\te$ affects the detection efficiencies  
in that shorter time scale events will, in general, 
be less densely sampled by monitoring teams than longer time scale events.  
Blending complicates
matters due to the ambiguity between light curves with different 
combinations of $\umin$ and blend fraction $f_B$ (Wo{\'z}niak \& Pacy{\'n}ski
1997), which can lead to ill-determined detection efficiencies if 
the blend fraction is poorly constrained (\S\ 7).  

The magnification patterns of 
close binaries ($b \ll 1$), wide binaries ($b\gg 1$), and binaries with
small mass ratios ($q\ll 1$), can be written as 
mathematical perturbations to the single lens pattern (\cite{martin1999}); 
the light curves they produce can be mistaken for those due to a single 
lens for a majority of source trajectories.  
The fractional deviation $\delta$, which is defined by 
\begin{equation}
\delta(t) \equiv { A(t)-A_0(t) \over A(t)}~,
\label{fracdev}
\end{equation}
where $A(t)$ is the binary-lens magnification and $A_0(t)$ is the
magnification of the best-fit single-lens model, quantifies 
the degree to which a binary-lens light curve deviates from a best-fit single 
lens model as a function of time.  

The best-fit single lens model need not have the same parameters $t_0$,
$\te$ and $\umin$ as the underlying binary and will vary depending
on the size and duration of the deviation, which is determined 
by the source trajectory $\theta$ and the binary-lens parameters $(b,q)$.  
The top panel of \fig{fig:defs} shows a light curve for 
binary system with $q=0.01$ and $b=1.0$, and two 
single-lens light curves: one assuming that the parameters $\te, t_0,
\umin$ are the same as in the underlying system, and the other 
the best-fit single lens curve.  
In this particular instance, the difference between the two
single-lens curves is not large, although this is not universally true
(see \S\ 5.1).  The bottom panel shows the fractional deviation of the
binary light curve from both single-lens light curves.  
In this example, the deviation is appreciable ($\delta>1\%$) 
for a large fraction of the light curve, 
but is large ($\delta > 5\%$) for only $\sim 5\days$, underscoring the 
need for high photometric precision in microlensing planet searches.   

\section{Detection Probabilities for Idealized Microlensing Planet Searches}

The success of microlensing survey and monitoring teams in 
implementing massive observational programs has prompted, 
and in part been driven by, theoretical work outlining 
optimal observational strategies, expected detection probabilities 
for binary lens perturbations, and methods for characterizing and  
extracting additional information about the companions. 
 
Using heuristic arguments based on rough scaling relations for 
the size of planetary caustics, Mao \& Paczy{\'n}ski (1991) estimated in their 
seminal paper that 3\% of microlensing light curves should cross 
caustics and thus show planetary deviations if all lens have a 
Jupiter-mass planet at instantaneous angular separations 
comparable to the Einstein ring radius of the primary.  
Randomizing the orbital phase and inclination of the orbital 
plane to obtain the distribution of angular separations $b \, \theta_E$ 
on the sky corresponding to a real orbital radius $a$,  
Gould \& Loeb (1992) estimated that planets with 
characteristics like our own Jupiter and Saturn orbiting 
solar-type stars halfway to the Galactic center 
would have detection probabilities of 17\% and 3\%, respectively. 
Their substantial probability is based on the assumption non-caustic 
crossing perturbations as small as $\delta \gsim 5\%$  
can be detected regardless of their duration.  
Performing a similar study, Bolatto \& Falco (1993) obtained 
a higher detection probability of $\sim$40\% for Jovian-like  
planets, requiring that the {\it integrated\/} difference between 
the binary light curve and a single-lens light curve of the same 
impact parameter $\umin$ exceed a threshold corresponding to 
a 10\% deviation over 36 hours.  
Bennett \& Rhie (1996) included for the first time the 
finite size of the source, enabling them to study planetary
systems with mass ratios as low as $10^{-5}$, comparable to the mass
ratio of the Earth and a parent star with mass $0.5 M_{\odot}$, whose 
caustics are smaller than the angular size of typical source stars.
They found a small but non-negligible detection probability 
of $\lsim 2\%$ for such low-mass planets, assuming a 50\% weather duty cycle 
for observations capable of detecting deviations as 
small as 4\% and as brief as $t_E/200$ ($\sim$2.4~hour).

All of these studies have concentrated on companions with instantaneous 
angular separations in the ``Lensing Zone,'' 
a region defined by $0.6 \le b \le 1.6$ in which planetary caustics 
may cross the source trajectory for events with $\umin < 1$, corresponding 
to an initial alerting magnification by the survey teams of $A > 1.34$. 
Recently, Di Stefano \& Scalzo (1999a, 1999b) have presented 
exhaustive studies of the probability of detecting planets in wide 
orbits that place gas giants like those in our own Solar System 
outside the Lensing Zone of typical lenses.  They conclude that 
such events may be detectable as isolated short duration events, 
which will alter the time scale distribution of all microlensing 
events, or as repeating events as the source trajectory passes 
first near the primary and then near the companion on the sky.  
The observing and analysis strategies of the survey 
and monitoring programs must be altered to optimally detect such planetary 
deviations, which, by definition, occur when 
the source is outside the Einstein ring of the primary.

Wambsganss (1997) has demonstrated the extraordinary
variety of perturbations that can be produced by given planetary systems 
depending on the trajectory of the source star.  
In principle, this variety might lead to ambiguity in the 
inferred properties of a lensing planet based on a measured light curve, 
a potentially serious difficulty for the characterization of detected 
planetary systems. 
Gaudi \& Gould (1997) found that potentially severe fitting ambiguities  
of a factor of $\sim$20 in the measured mass ratio could plague 
light curves with planetary anomalies, but that these 
could be alleviated by dense and accurate sampling 
of the planetary perturbations or simultaneous optical/infrared photometry.  
Gaudi (1998) notes that planetary microlensing anomalies may also 
be mistaken for particular types of binary source events; the 
ambiguity can be mitigated with similar additional observations.  
Griest \& Safizadeh (1998) showed that the planetary detection
probabilities for high magnification events can be quite large since 
the source necessarily passes close to the central caustic generated 
by any binary.  
Using several detection criteria, they found that 
Jupiter-mass planets can be detected with $\sim$100\% efficiency
in events with $\umin \lsim 0.1$ ($A_{\rm max} \gsim 10$) over 
a substantial range of parent-planet separations.  
Gaudi, Naber, \& Sackett (1998) pointed out that this 
necessarily implies that the presence of {\it multiple\/} planets would be 
revealed in high magnification events, since all companions 
contribute to the central caustic.  They also noted that   
unless the caustic morphology of higher multiplicity lensing systems 
is understood, fitting degeneracies will prevent the association of 
deviant light curves with well-characterized planets in such events. 

Such considerations have led to detailed proposals for 
observational strategies designed to maximize the 
detection of planets via microlensing.  For planets of all mass 
ratios, Peale (1997) has stressed 
the importance of a longitudinally-distributed network of southern 
telescopes for continuous monitoring, as has implemented by the 
PLANET collaboration since 1995 in its Jovian-mass search 
(\cite{albrow1997}, 1998).  Sackett (1997) points out that the high 
burden of proof demanded of non-repeating microlensing events 
necessitates a high-sampling rate for proper characterization of 
planetary anomalies; this suggests a wide-field imager at an 
excellent site as an ideal component in any microlensing search 
effort aimed at small mass planets.  
Both studies conclude that tens of planets per year may 
be detected by aggressive programs if every microlens has a planet 
in its Lensing Zone. 

\section{Defining Detection Criteria and Efficiencies for Actual Datasets}

Previous studies of detection probabilities of 
idealized microlensing planet searches (\S\ 3) differ in 
spirit and approach from the situation confronted by a researcher 
wishing to use an actual database of microlensing light curves to 
draw inferences about possible lensing companions.  First, 
real light curves are irregularly sampled, each with its own and 
usually varying photometric precision.  Second, only the post-alert portion 
of the light curve is available to current monitoring teams, resulting 
in differences in the monitored phase of each event.  Third, the intrinsic 
impact parameter distribution of the events is altered by choices made by 
both the discovery and monitoring teams; the actual distribution 
cannot be assumed to be uniform over any interval.  
Finally, because the true lensing parameters are unknown, a binary lens 
fit must be compared to the best-fit single lens model for a given 
light curve, not to a single-lens model in which the companion has been 
removed (see Fig.~\ref{fig:defs}).  Previous studies that did not use 
the best-fit single-lens model as the null hypothesis 
when calculating detection probabilities for idealized searches 
have overestimated the true detection probability (\S\ 5.1).  
These differences motivate the detection criteria and method of calculating 
detection efficiencies for actual datasets that we now describe.

All light curves contain information about the presence of companions
around Galactic lenses: obviously anomalous light curves signal the
possible presence of a companion while light curves without a detectable
anomaly signal the possible absence of (certain types of) companions.
With real datasets, both statements are probabilistic: the presence or
absence a particular type of lensing companion in a particular system
can only be made with a certain degree of confidence.  Ultimately, we seek
a method to characterize this probability consistently for all
light curves, whether obviously anomalous or not, so that the 
complete dataset can be used to constrain the distribution of planets
and other companions in orbit around Galactic lenses.

Because most microlensing light curves do not show obvious anomalies
indicative of companions, we begin here by quantifying the 
extent to which an apparently non-anomalous light curve can be taken 
as a sign that planetary companions of a given mass ratio
$q$ and projected separation $b$ are truly absent in the 
microlensing system under consideration.    
The absence of an observed planetary anomaly may be due to insufficient 
sampling, photometric precision, or monitoring duration, rather than the
absence of a planet.  
This means that the efficiency of detection 
must be computed separately for each event.  Intrinsic
parameters, such as source trajectory, source size, and blending, 
may also serve to `hide' the planetary anomaly from the observer, and 
must be disentangled from observational effects both to compute  
detection efficiencies of individual datasets and to formulate future 
observing strategies which will maximize the detection probability. 

Before we can begin to quantify the detection efficiency,
$\epsilon_i(b,q)$, of a lensing binary with microlensing parameters
$b,q$ using light curve $\ell_i$, we must define the meaning of
`detection.' Here, we will consider a planetary companion to a lens to
be `detected' in light curve $\ell_i$ if some combination of the
intrinsic binary parameters $b,q$ and $\theta$ produces a substantially
better fit (characterized by $\Delta\chi^2$) to the observed light curve
than the best-fit single lens model.  In making the comparison, the
parameters $t_0, \te, F_0, \umin$ and $f_B$ are allowed to vary to achieve
the best fit in both the binary and single lens models.  The meaning of
`substantially better fit' can be adjusted by altering the threshold
value of $\dchit$ that $\dchi= \chi^2_{\rm best\, single}- \chi^2_{\rm
best\, binary}$ must exceed.  

Since here we consider only light curves that are
not obviously discrepant from PSPL (by the criterion above), 
the angle $\theta$ between the source trajectory and binary axis 
is degenerate, and can be assumed to be drawn from a random distribution 
over the full range $0 <\theta \le 2\pi$. 
By {\it detection efficiency\/}, $\epsilon_i(b,q)$, we will mean the
probability that an actual planetary companion with mass ratio $q$ and
instantaneous angular separation $b$ would be `detected' by the
criterion above in a given {\it observed} light curve $\ell_i$, assuming 
a random source trajectory.  
Only a subset of source trajectory angles will produce detections 
by this criterion.  The efficiency incorporates both the intrinsic 
sensitivity of the observations to a given anomaly and the probability 
that such an anomaly occurs for random source trajectories. 
An efficiency $\epsilon_i(b,q)$ of zero implies that a lensing
companion with characteristics $b$ and $q$ would always escape detection
with these observations; $\epsilon_i(b,q)=1$ implies that the companions
would always be detected (if present) with data of this type and quality
regardless of source trajectory through the magnification pattern.
This efficiency can then be used to place a confidence level 
on the {\it non-detection} of $b,q$ binary anomalies in a given light curve.

Note that in order to draw inferences about the non-detection of planetary
companions of absolute mass $m_p$ (in $M_\odot$) and orbital radius $a$
(in AU), additional assumptions or information must be brought to bear
on the distribution of lens masses and distances.  Integrations over
orbital phase and inclination then must be performed to deproject the
instantaneous angular separation $b$ into a probabilistic distribution
function for orbital radius $a$.

\subsection{Qualitative effect of the Intrinsic Binary Parameters}

In addition to its dependence on observational parameters, 
the detection efficiency of a given light curve will be a 
strong function of the intrinsic mass ratio $q$ and 
instantaneous binary separation $b$ of the lens system.  

\begin{figure*}[t]
\epsscale{1.5}
\plotone{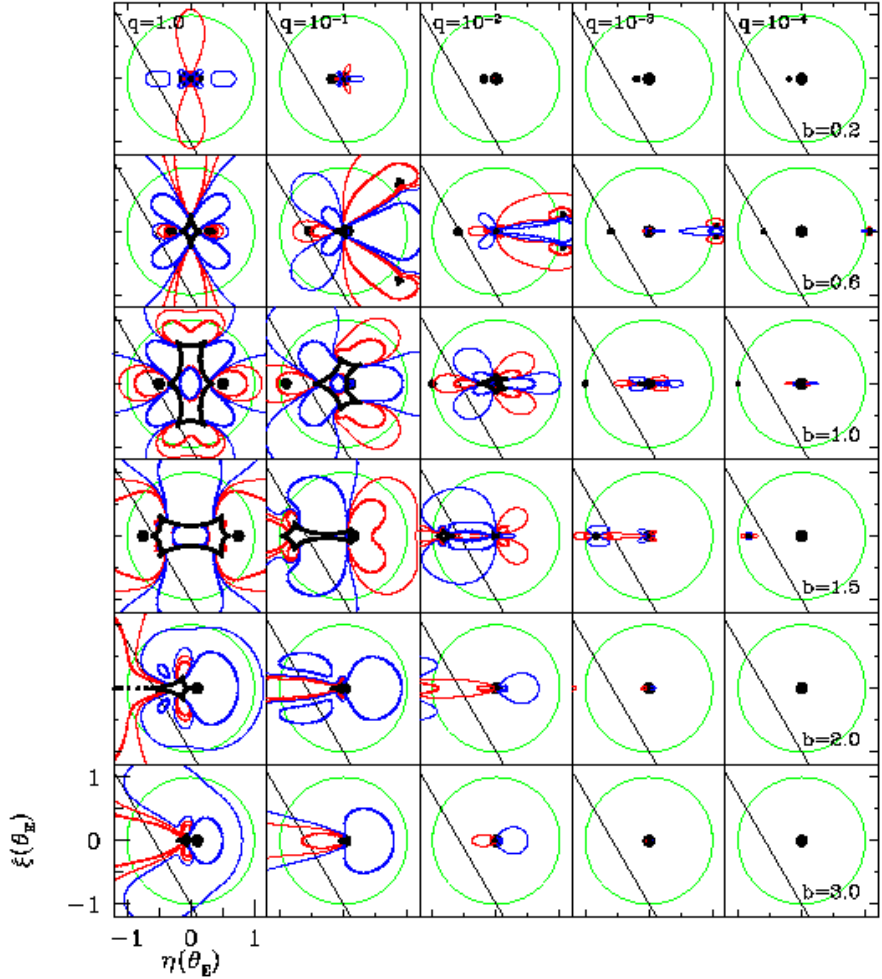}
\caption{
\footnotesize
Contours of fractional deviation $\delta$, for binary 
separations $b=0.2, 0.6, 1.0, 1.5, 2.0$, $3.0$ and mass ratios
$q=1.0, 10^{-1}, 10^{-2}, 10^{-3}$, $10^{-4}$, 
as a function of position in the source plane ($\eta,\xi$) 
in units of the Einstein ring radius of
the binary, $\theta_E$.
The red contours are $\delta=+0.01, +0.05$ (thin and heavy lines),
the blue contours are $\delta=-0.01, -0.05$ (thin and heavy lines), 
and the heavy black lines are the caustics, $\delta=\infty$.  
The green circle is the Einstein ring.  
Black dots mark the positions of the lensing masses.  
The thin straight line is a
trajectory with $u_{min}=0.5$ and $\theta=300$.   Light curves for
these trajectories are shown in Figure 3.}
\label{fig:contours}
\end{figure*}

\begin{figure*}[t]
\epsscale{1.5}
\plotone{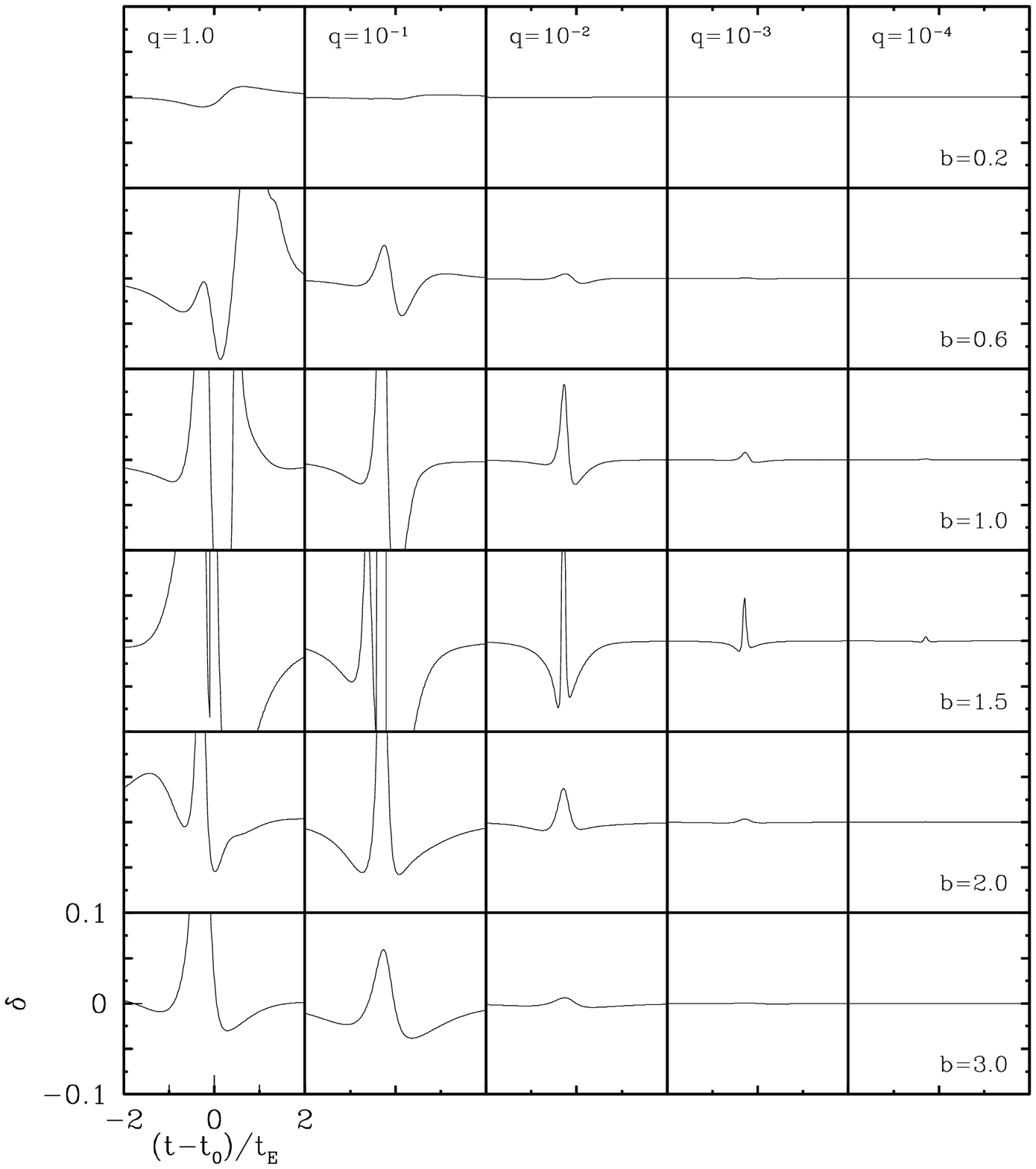}
\caption{
\footnotesize
The fractional deviation, $\delta$, from the single-lens light curve as
function of time from the peak of the single lens light curve, $t_0$, in
units of the Einstein ring crossing time $\te$, for the
trajectories shown in Figure 2. }
\label{fig:lcurves}
\end{figure*}

Three limits are of special interest.  First, as $b\rightarrow 0$, a binary
lens reduces to a single lens, so that most light curves will have 
small detection efficiencies to binaries with small separations.  
Second, for $b \rightarrow 1$, a binary lens
reduces to two isolated point lenses.  Only small $\umin$ trajectories
with $\theta\sim 0$ (i.e. parallel to the binary
axis) will betray the presence of both lenses (\cite{diandm1996});
most trajectories will produce light curves that are indistinguishable 
from single lens light curves.  Finally, for 
$q \rightarrow 0$, a binary lens reduces to a single lens.  Detection
efficiencies thus will be highest for $q\sim 1$ and $b\sim 1$, and will
decline for decreasing $q$ and outside an annulus centered near $b\sim 1$.

To illustrate these points, we calculate $\delta$ as a
function source position $(\xi,\eta)$ for a grid of $b$ and
$q$ values.  Since the best-fit single lens light curve, 
and therefore $\delta$, depends on the exact trajectory, 
we must adopt heuristic approximations in order to display 
these qualitative results in a single plot of $\delta$ contours. 
For $b \le 1.0$ and all $q$, we choose $\umin$ 
to be the minimum projected separation from the center-of-mass, 
and normalize $\te$ to the total mass of the binary.  
For $b > 1.0$ and $q\le 10^{-1}$, we choose $\umin$ to be the
minimum separation from the more massive lens component, and
normalize $\te$ to this component as well.  For $q=1$ and $b=1.5$ we
choose  $\umin$ to be the minimum projected separation from the 
center-of-mass, and normalize $\te$ to the total mass of the binary. 
Finally, for $q=1$ and $b\ge 2.0$, we choose $\umin$ to be the
minimum separation from one of the components, and normalize $\te$ to
the same component.  These choices were made in order
to approximate the transition between close and wide binaries and 
minimize the deviation {\it globally}.  Since this qualitative example 
does not use best-fit single lens parameters, $\delta$ will be
overestimated in all cases; we discuss this effect fully in \S\ 5.1.
To cover the full range of parameter space for which the
detection efficiency is high, we choose 5 values of $q$, logarithmically
spaced between $1.0$ and $10^{-4}$, and $b=0.2$, $0.6$, $1.0$, $1.5$,
$2.0$, and $3.0$.  The results are shown in \fig{fig:contours}, where 
contours of $\delta=\pm 0.01$ and $\pm 0.05$ are plotted with the
caustics ($\delta=\infty$).  Light curves are one-dimensional cuts through
these diagrams.  Since current survey programs rarely alert 
on-going events with $\umin \gsim 1$, only trajectories that pass 
within $\theta_E$ (circles in \fig{fig:contours}) of the primary lens 
will be observed by the monitoring teams.

In \fig{fig:lcurves}, light curves resulting from the sample trajectory in
\fig{fig:contours} are displayed.  This trajectory is rather typical, with 
$\umin=0.5$ and a randomly chosen angle $\theta$.
While for $q=1.0$ and $10^{-1}$, most trajectories will
exhibit considerable ($\delta \ge 0.05$) deviations, for $q \lsim
10^{-2}$, the magnitude of the deviation will depend strongly on the
value of the $\theta$.  For example, although the light curve in
\fig{fig:lcurves} for $q=10^{-3}$ and $b=0.6$ 
exhibits no significant deviation, inspection of \fig{fig:contours} 
reveals that a trajectory with the same impact parameter but $\theta \sim
90^{\circ}$ would be likely to have larger $\delta$.  
This illustrates why a calculation of the detection efficiency of 
a non-anomalous light curve must involve an integration over $\theta$, 
which is degenerate in the single-lens case.  
Figures \ref{fig:contours} and \ref{fig:lcurves} also illustrate 
why the detection efficiency of a light curve depends strongly 
on its $\umin$.  The sample trajectory for $q=10^{-3}$ and $b=1.0$, 
for example, does not deviate more than $\delta \sim 0.05$, yet  
almost all trajectories with smaller $\umin$ would exhibit much 
larger deviations.
 
Several conclusions can be drawn from inspection of Figures
\ref{fig:contours} and \ref{fig:lcurves}.
First, for $q\sim 1$ and $0.2 \lsim b \lsim 3.0$, nearly {\it{all}}
trajectories have deviations $\delta > 5\%$.  Second, for all separations 
$b$, events with $\umin \lsim 0.1$ will have a much higher detection
efficiency than larger $\umin$ events.  Third, for small mass ratios
($q\lsim 10^{-2}$), it is likely that only a small fraction of
detected events will exhibit caustic crossings, since, for these mass
ratios, the area covered by the caustics is considerably 
smaller than the area covered by the
$\delta=\pm 0.05$ contours.  Finally, for small $q$, detection
efficiencies for light curves with typical impact parameters 
will be substantial only for companions with separations 
$0.6 \lsim b \lsim 1.6$ (i.e., in the `Lensing Zone').

\begin{figure*}[t]
\epsscale{1.5}
\plotone{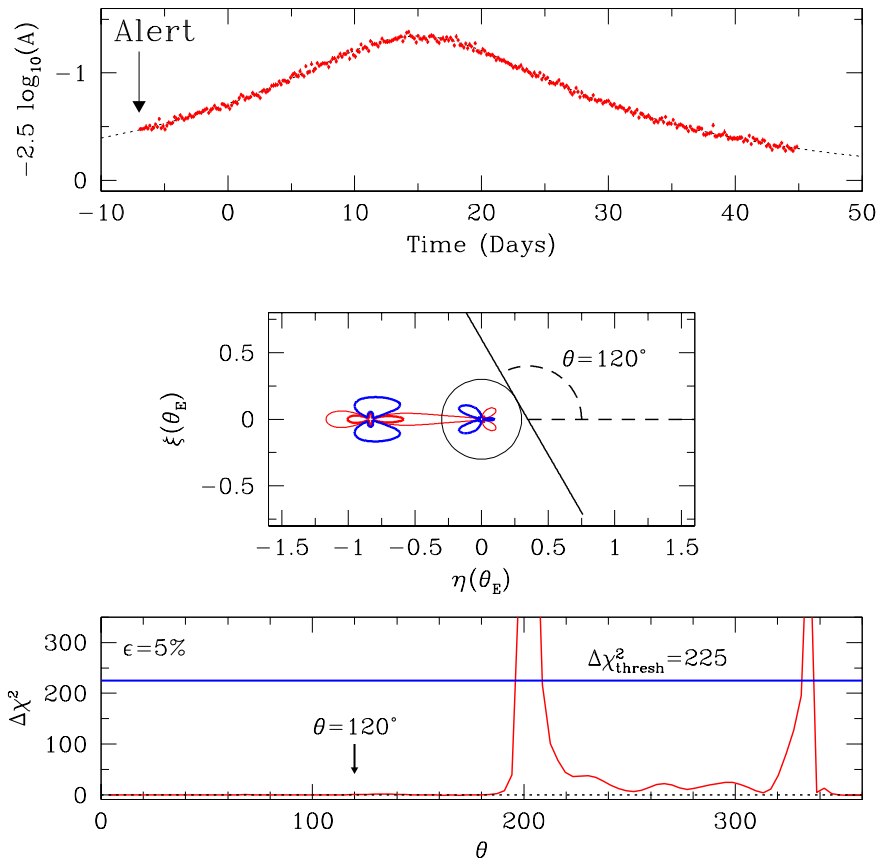}
\caption{
\footnotesize
Top panel: The difference in magnitude, $-2.5~\log_{10}(A)$, as a function of time for a simulated
microlensing event with photometric precision of $2\%$ and with uniform
sampling at the rate $\te/200$ from alert until $\te$. 
The dotted line shows the best-fit point-source
point-lens (PSPL) event, which has parameters $\te=15~{\rm days}, \umin=0.5,
t_0=15~{\rm days}$.  Middle panel:  Contours of constant fractional
deviation $\delta$ from the single-lens magnification, as a function of
source position $(\eta,\xi)$, in units of the Einstein ring radius 
$\theta_E$.  The solid line shows a trajectory with $\umin=0.5$ and
$\theta=120^{\circ}$.  The contours are the same as in Figure 2.  Bottom
panel:  The difference in $\chi^2$ between the best-fit binary-lens
light curve with $b=1.5$ and $q=10^{-3}$ and the best-fit PSPL
light curve, as a function of $\theta$.  The horizontal
line indicates a detection threshold of $\dchit=225$.  The detection
efficiency $\epsilon$ of the light curve in the top panel to 
companions with $b=1.5$ and $q=10^{-3}$ 
is the fraction of all possible trajectories for which
$\dchi>\dchit$. In this case, $\epsilon=5\%$. }
\label{fig:scheme}
\end{figure*}

\section{Description of the Algorithm}

An schematic of our basic algorithm is displayed in \fig{fig:scheme}.
The top panel shows a simulated light curve, which for the moment we use 
as a stand-in for an actual observed light curve.  (For details on
how this curve was generated, see \S\ 5.1.)  For simplicity, we assume
that the baseline flux is known perfectly ($F_0=1$) and that the event is not
blended ($f_B=0$).  We fit a restricted PSPL model with 
the three remaining free parameters $\te, \umin, t_0$ to this `observed'
light curve, and obtain $\chi^2_{\nu}=0.96$ for 
300 degrees of freedom, with best-fit parameters
($\te=15 \, {\rm days}, \umin=0.3, t_0=15.0 $).  This value of
$\chi^2_{\nu}$ indicates that the observed light curve is consistent with
the PSPL model.    In order to calculate the efficiency 
function $\epsilon(b,q)$ for this event, we 
must determine what fraction of all possible light curves arising from
a $b-q$ binary lens is {\it incompatible\/} with the observed light curve.  
The middle panel of \fig{fig:scheme} shows contours 
of constant $\delta$ for a binary
with $q=10^{-3}$ and $b=1.5$, and a sample trajectory with
$\umin=0.3$ and angle $\theta=120^{\circ}$ between the 
trajectory and the binary axis.  
Since this source path does not cross any regions of
significant deviation, the corresponding observed light curve would be 
consistent with PSPL within the precision of typical monitoring photometry.  
Light curves resulting from trajectories with other values of $\theta$  
would be inconsistent with the observed light curve.
In order to determine the incompatibility of a $b-q$ binary lens model 
with the observed light curve as function $\theta$, for each fixed $\theta$ 
we find the best-fit binary-lens light curve, but 
leaving $\umin, \te$ and $t_0$ as free parameters.  
We then calculate $\Delta\chi^2(b,q,\theta)$, 
the difference in $\chi^2$ between the
best-fit binary-lens light curve and the best-fit single lens light curve 
as a function of $\theta$.  
The bottom panel of \fig{fig:scheme} shows
$\Delta\chi^2$ as a function of $\theta$ for a binary with $q=10^{-3}$
and $b=1.5$.  Since only a small fraction of all possible trajectories would
give rise to binary-lens light curves that are statistically incompatible
with the data, the detection efficiency of this 
light curve is small for this parameter combination.  
Quantitatively, the detection efficiency is
simply the fraction of all possible trajectories ($0 \le \theta \le
2\pi$), for which $\dchi(b,q,\theta) > \dchit$.  Thus $\epsilon(b,q)$ is
given by,
\begin{equation}
\epsilon(b,q)\equiv{1 \over 2\pi} \int_0^{2\pi} {\rm d}\theta\,\, \Theta[\dchi(b,q,\theta)
- \dchit],
\label{ebq}
\end{equation}
where $\Theta[x]$ is a step function.  For the event depicted in \fig{fig:scheme},
$\epsilon=5\%$ for $b=1.5$ and $q=0.001$ and a detection criterion of
$\dchit=225$.  This process must then be
repeated for all $q$ and $b$, and then for all light curves that are
consistent with the point-lens model.

For Gaussian errors, $\chi^2$ is the best measure of
goodness-of-fit, and the significance of the detection can be altered by
adjusting $\dchit$, the minimum $\dchi$ between the best-fit 
single and binary lens models required for a detection. 
The choice of $\dchit$ required for a `detection' is arbitrary, but
it should be kept in mind that error distributions for actual monitored
events are far from Gaussian, and usually contain systematic
errors with unrecognized correlations at the few percent level.  
In light of this,  we choose a rather
conservative detection criterion, $\dchit > 100$.  
The choice of the appropriate detection
criterion for realistic error distributions will likely depend
sensitively on, and be determined by, the actual error distributions
themselves.  We return to this point with a discussion of the
dependence of detection efficiencies on the detection threshold in \S\ 5.1.

The basic steps in calculating the detection efficiency of
events consistent with the PSPL to a specific $b-q$ binary are 
summarized below.  With the appropriate modifications, a similar 
algorithm can be applied to the analysis of
detection efficiency for {\it any} microlensing anomaly, including
those due to binary sources, lens rotation and parallax effects.

\begin{description}
\item[{(1)}] Fit each event with a single lens model by
minimizing $\chi^2$ (or some other suitable goodness-of-fit estimator).
Evaluate $\chi^2$ for this model.
\item[{(2)}] Hold the angular separation and mass ratio $(b,q)$
fixed. For each source trajectory $\theta$, find the binary lens model
that best fits the observed light curve, leaving $\te, t_0,$ and $\umin$
as free parameters. Evaluate the difference $\dchi (b, q, \theta)$ between the
single-lens and binary-lens fits.
\item[{(3)}] Find the fraction of all binary-lens fits for the given $(b,q)$ that satisfy the detection criterion (e.g. $\dchi > \dchit$).  This is the
detection efficiency $\epsilon(b,q)$ for this event for the assumed
separation and mass ratio.
\item[{(4)}] Repeat items (2) and (3) for all $(b,q)$.  
This gives the detection efficiency for the $i$th event 
as a function of $b$ and $q$, $\epsilon_i(b,q)$.
\item[{(5)}] Repeat items (1)-(4) for all events.
\end{description}

The steps itemized above assume that the baseline flux $F_0$ is known
perfectly, and that the event is not blended, ($f_B=0)$.  In reality, one
must always fit for the baseline flux $F_0$ and blend parameter $f_B$.  
If an event is truly blended, including $F_0$ and $f_B$ in the fitting 
procedures can have a strong effect on the computed detection efficiency 
of the resulting light curve. Similarly, the algorithmic 
outline above assumes that the source can be treated as point-like.  
Including finite source sizes can also have a significant
effect on the inferred detection efficiency of a given event.  
In order to obtain an accurate estimate of 
the detection efficiency, these effects must be included in the 
fitting procedure, and either the light curve itself or other data 
used to constrain the blending and finite source size parameters. 
In order to clearly delineate the effects of blending, finite source 
size, and the choice of the detection criterion, we will first assume 
that the baseline flux is known perfectly, the blending is negligible, 
and the source can be approximated as a point-like.  The effects of 
detection criterion, finite source size and blending on the detection 
efficiency $\epsilon(b,q)$ are then explored 
separately in \S\S\ 5.1, 6 and 7, respectively.

The detection efficiencies calculated in the prescribed way for
non-anomalous events can be used in several ways: (1) to
place quantitative constraints on the absence of planets of certain
$b,q$ in non-anomalous lensing event; (2) to estimate the average
detection efficiency $\epsilon(b,q)$ for a given dataset; (3) to
estimate $\epsilon(b,q)$ for hypothetical datasets as a guide to future
observational programs, and (4) as a proxy for the detection efficiency
of observed anomalous events, for which additional challenges exist
(see discussion in \S\ 8).

\begin{figure*}[t]
\epsscale{1.5}
\plotone{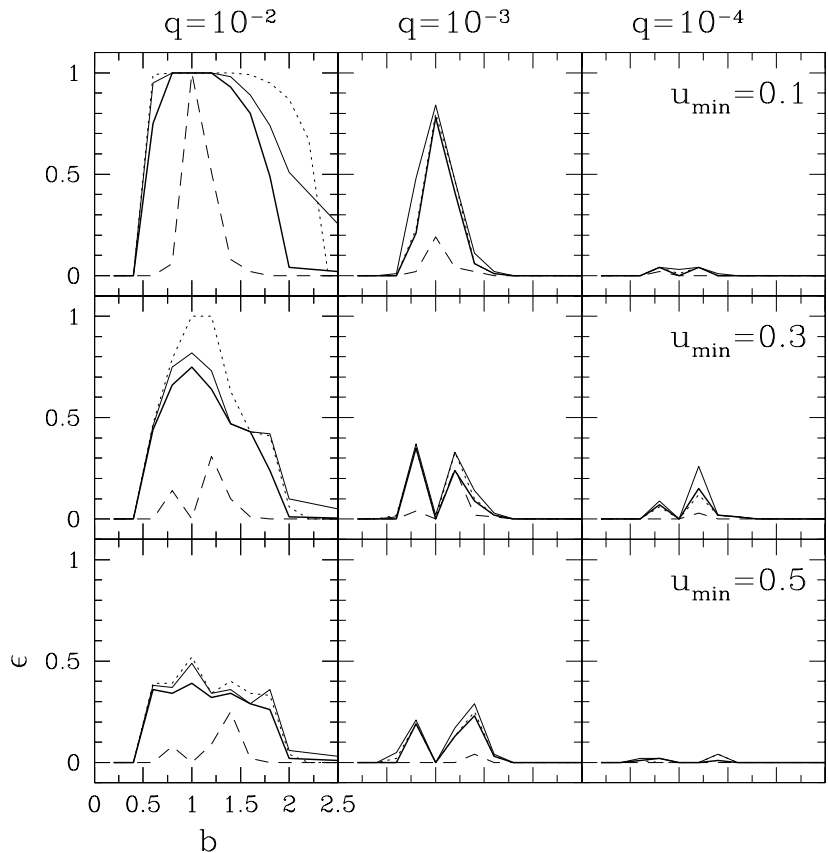}
\caption{
\footnotesize
The detection efficiency, $\epsilon(b,q)$ as a function of the instantaneous
separation of the binary, $b$ for three different mass ratios  
($q=10^{-2}, 10^{-3},$ and $10^{-4}$) and impact parameters 
($\umin=0.1, 0.3,$ and $0.5$). Three detection criteria are  
shown: $\Delta\chi^2 > 225$ (thick solid line), $\Delta\chi^2 > 100$ 
(thin solid line), and caustic crossing (dashed line). 
The efficiency $\epsilon$ using $\Delta\chi^2 > 225$, but without fitting the
binary-lens light curve, is also shown (dotted line).  
Throughout it is assumed that the events are sampled uniformly with $2\%$ 
photometry at the rate $\te/200$ from alert to $\te$.  See \S\ 5.1.}
\label{fig:pseps}
\end{figure*}

\subsection{Application to Artificial Data}

In order to explore more fully the effect of the parameters $\umin$,
$b$, $q$, and detection criterion 
on the detection efficiency, and to test the robustness of
the algorithm, we generate artificial light curves and
calculate their detection efficiency $\epsilon (b,q)$.

Each simulated event is assumed to be alerted at $A =1.54$ (the smallest
amplification that the MACHO team will alert, \cite{alcock1997b}), and
then continuously observed at uniform intervals of $\te/200$ until
either $\te$ after the peak or $3\te$ after the peak.  A more realistic
light curve would contain gaps due to bad weather or other observing
conditions. In order to isolate as much as possible the intrinsic
dependencies of the detection efficiency, we use uniform sampling, and also   
assume here $f_B=0$ and $F_0=1$.
At each observation, a residual is drawn from a Gaussian distribution with 
$\sigma=\sigma_0 A_0$, where $\sigma_0=1\%$ or $2\%$, and $A_0$ is given
by Equation (\ref{mag0}).  These parameters are roughly consistent with 
the best sampling and photometric accuracy of the PLANET collaboration
(\cite{albrow1998}).  Hereafter, all results will be for 
light curves with $\sigma_0=2\%$ and observations until $\te$ unless
otherwise noted.  Three values of the impact
parameter, $\umin = 0.5$, $0.3$, and $0.1$, are investigated.  

For each light curve, we calculate $\epsilon (b,q)$ using steps
(2-4) in \S\ 5.  When fitting the binary, we employ a downhill-simplex
method (\cite{cookbook}), which usually converges quickly and robustly to the minimum.  The fitted single-lens parameters are used as an initial guess, 
since best-fit binary parameters are typically close to the single-lens values
for small mass ratios $q$.  

Since the fitting procedure can be
computationally expensive, we sample $\theta$ only at intervals of $2\pi/100$; 
our efficiencies are thus limited to a resolution of $\Delta\epsilon=0.01$.
Although one would like to sample the $(b,q)$ plane as densely as
possible, we are again limited by computational expense.  Since
planetary events are the primary interest of most monitoring
collaborations, and nearly equal mass binaries ($10^{-1} \le q \le 1.0)$
will have $\epsilon \sim 100\%$ in the Lensing Zone anyway, we choose to
restrict our attention to $q \le 10^{-2}$.  Furthermore, the effect of
finite source sizes and blending, which we wish to investigate, will be
substantially less dramatic for $q \gsim 10^{-1}$ than for lower
mass-ratio systems.  We choose three mass ratios: $q=10^{-2}$, $10^{-3}$
and $10^{-4}$.  Using a different detection algorithm, 
Bennett \& Rhie (1996) found detection
probabilities of $\sim$2\% for $q=10^{-5}$.  Since this is comparable
to our resolution $\Delta \epsilon=0.01$, we will not extend our
analysis to mass ratios smaller than $q=10^{-4}$.  We calculate $\epsilon$
from  $b=0.2$ to $2.0$ at intervals of $0.2$, and then again at $b=3.0$.

Previous explorations of planetary microlensing detection probabilities
have {\it not} used the best-fit binary lens light curve, which is
computationally expensive to compute.  Holding $\te$, $t_0$ and
$\umin$ fixed will cause one to overestimate the detection efficiency in
two ways.  First, for large mass ratios $q\gsim 10^{-2}$, the secondary
cannot simply be treated as a perturbation to the primary light curve.
The presence of the secondary will have a significant effect on the
global (averaged over all trajectories) values of the best-fit
parameters.  For example, a close ($b\ll 1$) equal mass binary will
have a time scale that is a factor of $2^{1/2}$ larger than an otherwise identical 
wide ($b \gg 1$) binary.  
Second, for small mass ratios $q \lsim 10^{-3}$, the majority of the
detection efficiency will arise from relatively small, $\delta \lsim
10\%$, deviations.  These deviations can be suppressed below the detection
criterion if the fit is allowed to adjust to compensate for them (\cite{gands1998}).
In order to facilitate comparison
between our results and previous calculations, and to gauge the error
induced by holding the parameters $\te$, $t_0$, and $\umin$ fixed when
calculating the difference between binary-lens and
single-lens magnifications, we have also calculated $\epsilon (b,q)$ without
finding the best-fit binary, assuming instead that $\te$, $t_0$ and 
$\umin$ are the same for the binary and single lens fits.  
For $b<1$, we
choose the center-of-mass as the origin of the binary; for $b>1$, we
choose the position of the primary.

\begin{table*}
\begin{center}
\begin{tabular}{|c||c||c|c|c|c|}
\tableline
	$u_{min}$ &q &$\dchit=100$ &$\dchit=225$ &C.C. &$\dchit=225$\\
                  &  &             &             &     &  w/o fit\\
\tableline
\tableline
0.1 & 0.01 & $98.0\%$  & 94.1 & 33.2  & 99.8 \\ 
   & 0.001 & 38.3      & 29.3 & 5.4   & 31.3 \\ 
   & 0.0001& 4.6       & 1.6  & 0.4   & 1.8  \\
\tableline
\tableline
0.3 & 0.01 & 64.3   & 59.1 & 11.1 & 77.4   \\
    & 0.001 & 17.6  & 13.8 & 6.3  & 16.0   \\
    & 0.0001& 7.5   & 4.9  & 0.6  & 4.1    \\
\tableline
\tableline
0.5 & 0.01  & 37.9  & 34.3 & 8.6  & 40.3   \\
    & 0.001 & 14.3  & 11.3 & 0.8  & 11.9   \\
    & 0.0001& 1.4   & 0.7  & 0.0  & 0.9    \\
\tableline
\end{tabular}
\end{center}
\tablenum{1}
{\bf Table 1} Lensing Zone Detection Efficiencies $\elz (q)$:
Point Source \label{tbl-1}
\end{table*}

In \fig{fig:pseps} the detection efficiency $\epsilon(b,q)$  
is displayed as a function of the
dimensionless angular separation $b$ for events that are followed
until $t=\te$.  Three different impact
parameters ($\umin=0.1, 0.3,$ and $0.5$) and mass ratios ($q=10^{-2},
10^{-3},$ and $10^{-4}$) are investigated using different detection criteria: $\dchit=100$, $\dchit=225$, and the criterion that the trajectory must cross a
caustic in order to be detected.  Also shown is $\epsilon(b,q)$ for
$\dchit=225$ without fitting the binary light curve.  For all events,
we have assumed a photometric accuracy of $2\%$ and that observations
are carried out from alert until $1\te$ after the peak.

\fig{fig:pseps} illustrates several points.   
First, the difference between the
detection efficiency calculated using the two different thresholds,
$\dchit=225$ and $100$, is small and approximately 
constant at $\Delta\epsilon \sim 0.05$ over most of the parameter 
space considered.
However, since the magnitude of $\epsilon(b,q)$ decreases with
decreasing mass ratio $q$, the fractional difference increases. 
The exact choice of $\dchit$ thus has little effect on $\epsilon(b,q)$ 
for deviations well above the detection threshold.  
For perturbations near the detection limit, however, such as those arising from
companions with mass ratio $q=10^{-4}$, the detection efficiency 
can vary by a factor of two depending on the choice of detection criterion. 
Second, the error induced by not using the best-fit binary lens
light curve can be substantial for $q=10^{-2}$, because 
such companions cause significant anomalies over a large fraction of the light
curve so that the parameters of the best fit binary- and single-lens
models can differ dramatically. This is especially true for high
magnification events with $b > 1$, for which the fractional deviation $\delta$ 
from the best single-lens model depends critically on the choice of the 
binary origin, which can be significantly different from the primary lens
position depending on the value of $b$.  The error
induced is also large for $q=10^{-4}$, because these
deviations are very near the detection limit.  For $q=10^{-3}$, the
deviations caused by the companion can be treated as a perturbation to
the primary light curve, and are well above the detection
limit, minimizing the effect of not using the best-fit binary; 
we caution, however, that this is unlikely to be true for 
all realizations of realistic sampling.  We conclude that one can
avoid fitting the binary lens light curve for mass ratios $q\lsim
10^{-3}$ only if the deviations are well above the detection limit of
the observed light curve. Finally, as noted by Gould \& Loeb
(1992), we find that caustic crossing events are likely to comprise only
a small fraction of all detected events.  This is especially important
because non-caustic crossing events are more prone to degeneracies and
thus the most difficult to characterize (\cite{gandg1997}; \cite{me1998}).

In order to quantify the effects of the various detection criteria
and impact parameters on $\epsilon(b, q)$, we tabulate in 
Table~\ref{tbl-1} average
detection efficiencies $\elz(q)$ for the curves in \fig{fig:pseps} 
integrated over the Lensing Zone (where
the detection efficiency is the highest), $0.6 \le b \le 1.6$,
\begin{equation}
\elz (q) \equiv \int_{0.6}^{1.6}{\rm d}b \,  \epsilon (b,q)~. 
\label{elenszone}
\end{equation}
The accuracy of these results are limited by the fact that we 
sample $\epsilon(b, q)$ only at intervals of $\Delta b = 0.2$ in
this zone.  For any given event, however, the results are more secure, so 
that comparisons between $\elz$ for $(\umin=0.1, q=10^{-2})$ 
and $(\umin=0.1, q=10^{-3})$
should be more reliable than comparisons between $(\umin=0.1, q=10^{-2})$ 
and $(\umin=0.3, q=10^{-2})$.  
Table~\ref{tbl-1} illustrates that the
fractional error induced by not fitting the binary light curve is
smallest for $q=10^{-3}$, $\lsim 15 \%$.  For $q=10^{-2}$ and $q=10^{-4}$ however, the error can be 
considerably larger, $\sim 20\%$ for $\umin=0.3$ and $q=10^{-2}$, 
and $\sim 30\%$ for $\umin=0.5$ and $q=10^{-4}$.  
The fractional difference in $\elz$ between $\dchit=100$
and $225$ can be substantial, especially for $q=10^{-4}$, where it is
always greater than $50\%$. 
Finally, caustic crossing anomalies comprise
a relatively small fraction of the total events, representing at most
$35\%$ of the integrated detection efficiency, and decrease in importance
for large impact parameters and smaller mass ratio.

If all lensing primaries have planets distributed uniformly in $b$, 
the numbers in Table~\ref{tbl-1} represent the fraction
of all events with the given $\umin$ that would exhibit detectable 
deviations with the given $q$ and $0.6 \le b \le 1.6$.  
For larger mass ratio companions with $q=10^{-2}$, 
the detectable fraction is quite large, $\elz (q) \gsim 30\%$, 
and remains substantial, $\elz (q) \gsim 10\%$, 
even for ``Jovian'' companions with $q=10^{-3}$.  
For small companions with $q=10^{-3}$, however, the detectable lensing 
zone fraction drops significantly below $10\%$ in all cases, 
although the exact numbers are somewhat uncertain due
to the poor sampling in $b$.  We conclude that microlensing will be able
to place strong constraints on the frequency of double lenses with $q \gsim
10^{-2}$ and mild constraints on systems with $q=10^{-3}$ companions, 
but will be unable to meaningfully constrain systems with $q\lsim 10^{-4}$, 
unless the sampling and photometric precision are significantly better 
than those assumed in these simulations.

\begin{figure*}[t]
\epsscale{1.5}
\plotone{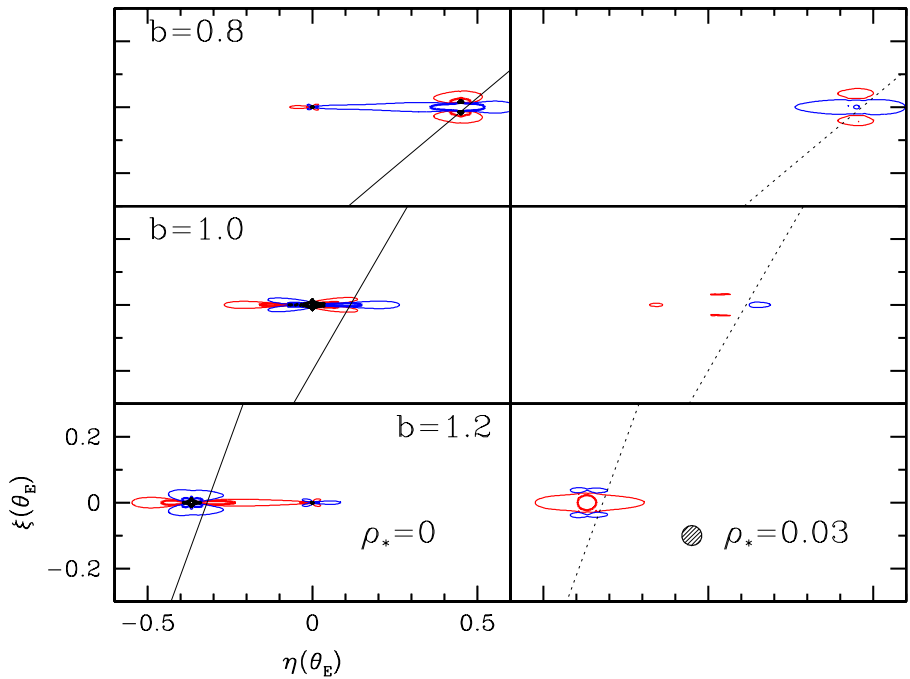}
\caption{
\footnotesize
Contours of constant fractional deviation, $\delta$, from the
single-lens magnification for a point-source (left
panels) and source with radius $\rho_*=0.03$ (right panels), as a function of
source position $(\eta,\xi)$, in units of the Einstein ring radius,
$\theta_E$, for mass ratio $q=10^{-4}$ and 
dimensionless separations $b=0.8, 1.0, 1.2$.  The
contours are the same as in Figure 2.  The
light curves arising from the sample trajectories (solid and dotted
straight lines) are shown in Figure 7.  }
\label{fig:fscontours}
\end{figure*}

\begin{figure*}[t]
\epsscale{1.5}
\plotone{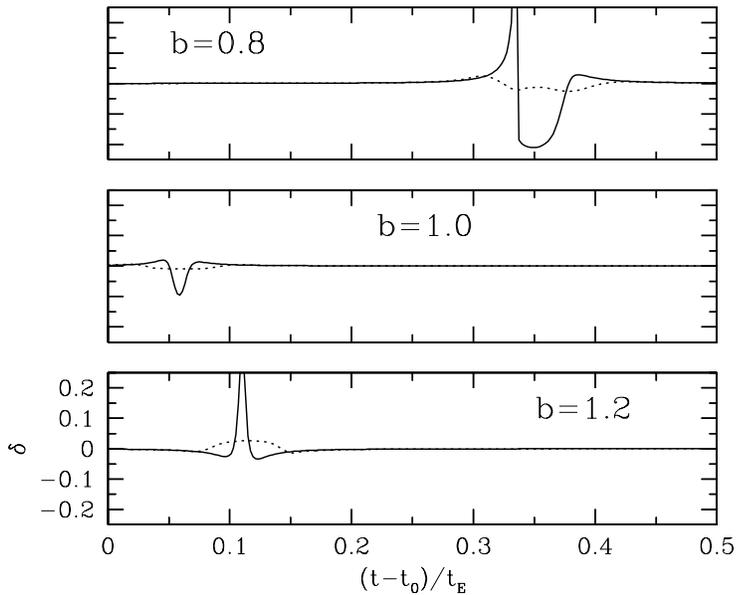}
\caption{
\footnotesize
The fractional deviation, $\delta$, from the single-lens light curve as
function of time from the peak of the single lens light curve, $t_0$, in
units of the Einstein ring crossing time $\te$, for the
trajectories shown in Figure 7.  The solid line is for a point-source,
and the dotted lines are for a source of radius $\rho_*=0.03$.}
\label{fig:fslcurves}
\end{figure*}

\begin{figure*}[t]
\epsscale{1.5}
\plotone{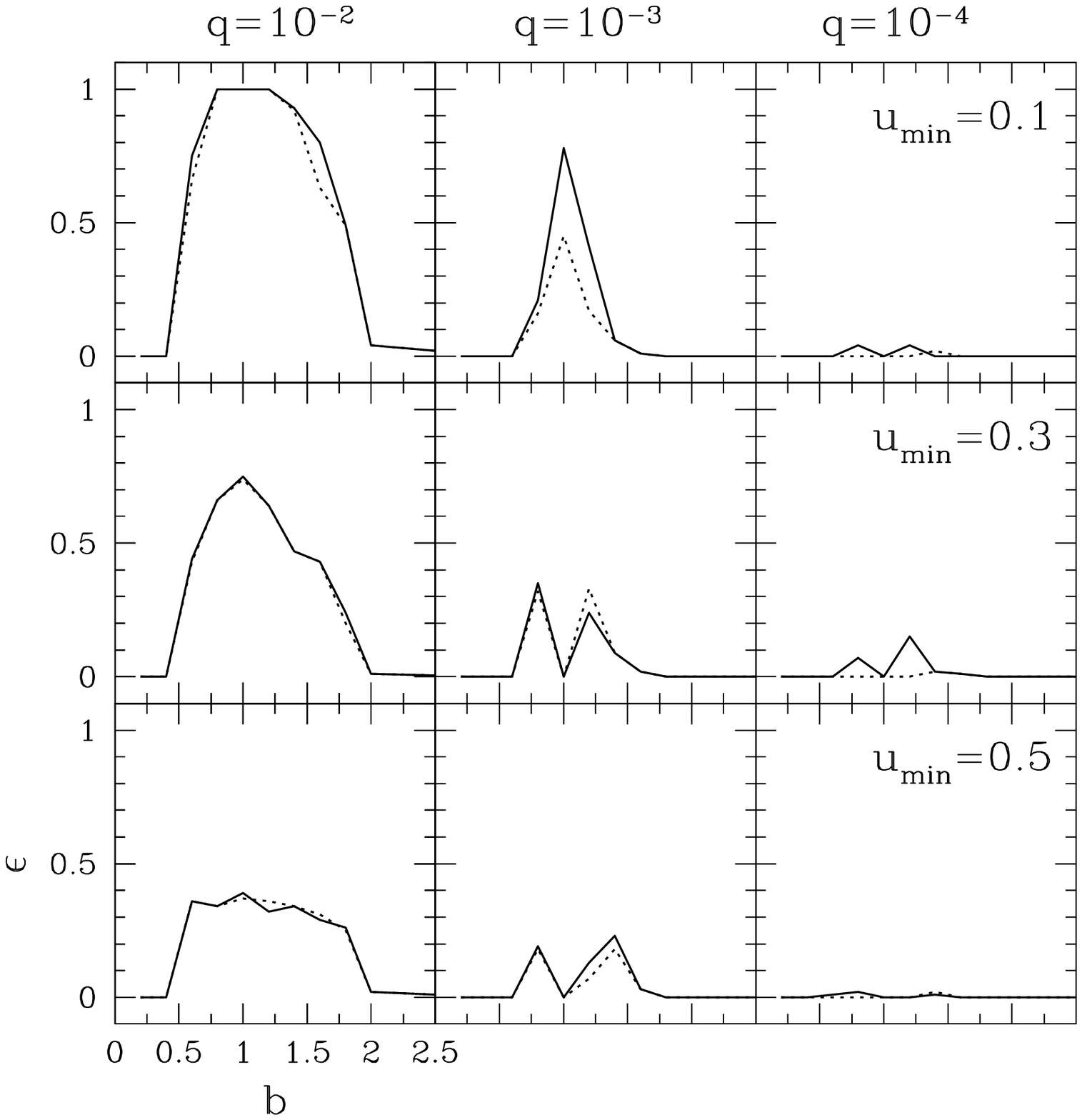}
\caption{
\footnotesize
The detection efficiency $\epsilon$ as a function of the
dimensionless separation of the binary, $b$, for three different mass 
ratios $q$ and three events with different minimum impact parameter $\umin$.
The solid line is $\epsilon$ for a point-source and a detection 
criterion $\Delta\chi^2 > 225$; the dotted line is $\epsilon$ using the
same detection criterion for a source of radius $\rho_*=0.03$.
It is assumed that the events are sampled uniformly with $2\%$ 
photometry at the rate $\te/200$ from alert to $\te$. }
\label{fig:fseps}
\end{figure*}

\section{Finite Source Effects}

The calculations in \S\ 5 implicitly assumed that the microlensed 
source was point-like, so that the magnification of the source 
is infinite at the caustics.  The magnification $A_{\rm fs}$ of a source 
with finite size is given by the integral of the
point-source magnification over the face of the star, 
\begin{equation}
A_{\rm fs}(t)= {{\int {\rm d}^2 r\, A(t;{\bf{r}})I({\bf{r}})}\over {\int
{\rm d}^2 r\,I({\bf{r}})}}~,
\label{fsmag}
\end{equation}
and is equivalent to the intensity-weighted area of
the images (numerator) divided by the intensity-weighted area of the
unlensed source (denominator).  Here $I({\bf{r}})$ is the intensity 
profile of the source.  The finite size of the source smoothes
and broadens the discontinuous jumps in magnification near caustics by
an amount that depends on the angular size of the source, $\theta_*$, in
units of $\theta_E$,
\begin{equation}
\rho_* \equiv {\theta_* \over \theta_E} = {R_* \over R_E}{\dol \over
\dos}= 0.03 \left({M \over 0.2 M_{\odot}}\right)^{1/2} \left(R_*  \over 10 R_{\odot}\right)~,
\label{rhostar}
\end{equation}
where $R_*$ is the physical size of the star.  For the scaling
relation on the right of \eq{\ref{rhostar}}, 
a lens distance of $\dol=6 \kpc$ and a source
distance of $\dos=8 \kpc$ has been assumed.  For a given $\rho_*$,
uniform sources ($I={\rm constant}$) will have a larger effect on the
magnification than limb-darkened sources.  Since we are interested 
primarily in the magnitude of the effect that finite sources will have on the
detection efficiency, we will assume a (less-realistic) uniform source
profile, which also increases computational speed.

Since the caustic of a point lens is a single point at the center of the
lens ($u=0$), the magnification of a finite source will differ 
from that of a point source only when the source approaches the
center of the lens, $\umin \sim \rho_*$.  For typical sources and
lenses, $\rho_* \ll 1$ (c.f. \eq{\ref{rhostar}}), so that for single 
lenses finite source effects are noticeable only 
in high magnification events
($\umin \ll 1$), for which $A\sim u^{-1}$ (\eq{\ref{mag0}}).  
Assuming a uniform source, the single lens magnification can be found
analytically in this limit (\cite{bible}),
\begin{equation}
A_{\rm fs,0}(t)=A_0(t){\cal{B}}[z(t)],
\label{fsmag0}
\end{equation}
where
\begin{equation}
{\cal{B}}(z)=\cases{ {4\over \pi}E(z) & $z \le 1$\cr {4\over \pi}z
\left[ E(1/z)- (1- z^{-2}) K(1/z) \right] & $ z\ge 1$\cr }~.
\label{calb}
\end{equation}
Here $K$ and $E$ are the complete elliptic integrals of the first and
second kind, and $z\equiv u/\rho_*$.

For a binary lens, the finite size of the source affects the
magnification whenever the source approaches a caustic or, 
more precisely, wherever the second derivative of the magnification is
large.  As can be seen in \fig{fig:contours}, for binaries with $q\sim 1$  
much of the region inside the Einstein ring satisfies this condition:  
caustic approaches and crossings will be common.  
On the other hand, the results of \S\ 5.1 
indicate that caustic crossings make up only a small fraction of all
detectable events for binaries with mass ratios consistent
with planetary systems ($q \lsim 10^{-2}$).  
Nevertheless, as can be seen from \fig{fig:contours}, in order 
to produce a detectable event trajectories must pass close to caustics, where 
the gradient of the magnification is large and finite source effects 
are non-negligible. 
Finite source magnifications for binary lenses 
cannot be found analytically.  Numerical integration of the point source
magnification over the face of the star is difficult, as the divergent
magnification near caustics causes the results to depend critically on
the integration grid size.  A more robust method is to compute 
the total area of all images and then divide by this area by that of 
the source to find the magnification of the finite source at that position.
Numerous methods have been suggested; we will 
integrate over the boundary of the images (\cite{kands1988}; 
\cite{gandgauch1997}; \cite{martin1998}).  For alternative methods, see
Bennett \& Rhie (1996), Wambsganss (1997), and Griest \& Safizadeh (1998).

What is relevant to this discussion is not the difference between the
finite source and point source magnification, but the effects
of finite source size on the determination of the detection efficiency
$\epsilon$.  For nearly equal-mass binaries $(q \gsim 10^{-1})$, the
magnification may be altered considerably by finite source 
effects without substantially altering $\epsilon$.  
This can be seen by comparing the size of the deviations ($\delta$)  
in the $q=1.0$ and $q=10^{-1}$ panels of \fig{fig:contours} to the size 
of a large source ($\rho_* =0.03$).  For mass ratios as 
small as $q \lsim 10^{-3}$, however, the size of the $\delta$ structures 
is comparable to that of a large source, and finite source 
effects are important to a proper determination of the 
detection efficiency $\epsilon$.  
Roughly speaking, finite source effects become important 
whenever the source size becomes comparable to
the Einstein ring of the companion, $\theta_* \gsim \theta_p = q
^{1/2}\theta_E$, or,
\begin{equation}
\rho_* \ga q^{1/2}.
\label{fslimit}
\end{equation}
This criterion is satisfied at $\rho_* \simeq 0.1$ for $q \simeq
10^{-2}$, whereas for $q \sim 10^{-3}$ source sizes $\rho_* \gsim 0.03$ 
will begin to seriously affect $\epsilon$.  Since the largest sources
routinely monitored in the Galactic bulge are clump giants, 
with $\rho \simeq 0.03$,  finite source effects will be
negligible for $q > 10^{-3}$, but must be considered for smaller mass
ratios.

Although the magnitude of the perturbation will always be suppressed in
the presence of a finite source, it will also be broadened.  Finite 
sources thus have competing effects on the detection efficiency
$\epsilon$: $\epsilon$ is decreased because previously significant
deviations are suppressed below the detection threshold, but increased 
for those trajectories for which the limb of the star grazes
a caustic (or high magnification area) yielding a significant deviation
where no significant deviation would have occurred for a point source.  
The net result of these two competing effects will depend on
the specific value of $q$ and $b$.

Contours of constant fractional deviation $\delta$ of a binary from a
single lens magnification, are illustrated in \fig{fig:fscontours} 
for both a point source and a finite source.
We choose $\rho_*=0.03$, a relatively large source, and $q=10^{-4}$, the
smallest mass ratio we consider, in order to present a scenario in which
the source size will have an extreme effect.  For the finite-source
cases, the fractional deviation is computed with respect to a {\it
finite-source} point-lens magnification, as given by \eq{\ref{fsmag0}}.
\fig{fig:fscontours}  clearly demonstrates that, 
for this mass ratio and source size,
the differences between the point and finite source magnification are
dramatic.  Both the shape and size of the $\delta$-contours are altered
considerably.  

As can be seen from \fig{fig:fscontours}, 
planetary perturbations with $b < 1$ are qualitatively different 
than those with $b>1$ (Gould \& Loeb 1992; Gaudi \& Gould 1997; 
Wambsganss 1997).  Consider the
case $b=0.8$: the perturbation is substantially depressed by the finite
source and the $\delta=5\%$ contours have nearly disappeared.  This is
because, for $b<1$, regions of constant positive and negative
deviation are closely spaced and of nearly equal area so that the 
smoothing induced by a large source tends to cause a cancellation 
leaving a deviation that is nearly zero (\cite{bandr1996};
\cite{gandgauch1997}).   The effect is even more prominent for $b\sim 1$,
where the regions of positive and negative deviation are especially 
closely spaced (Bennett \& Rhie 1996). Obviously, for these two
parameter combinations, the planet is unlikely to be detected and
$\epsilon\rightarrow 0$. For $b>1$, regions of positive deviation
encompass considerably more area than those of negative deviation (at 
a fixed value of $|\delta| \gsim 5\%$), and the cancellation is less
dramatic.  As a result, $\epsilon (b>1)$ will be less affected by finite
sources than $\epsilon(b\le 1)$. In both cases, the perturbations
caused by the central caustic (near $\zeta=0, \eta=0$) have dropped
below $\delta=1\%$.  Central caustics are an important channel to 
planet detection in high magnification events; 
detection efficiencies for these events
will be highly sensitive to $\rho_*$ (\cite{gands1998}).
Fortunately, this is a class of events for which $\rho_*$ can often be
measured.

In \fig{fig:fslcurves}, the light curves resulting from the trajectories
shown in \fig{fig:fscontours} are displayed.  
The trajectories were chosen to create a 
significant point source fractional deviation, but are otherwise
representative.  Dramatic cancellation can be seen in the
light curves for $b=0.8$ and $b=1.0$.  For $b=1.2$, the deviation is
substantially suppressed but is also broader.  For photometry of 
sufficient precision, the detection efficiency $\epsilon$ for this
parameter combination will actually be {\it increased}.

\bigskip

\begin{center}
\begin{tabular}{|c||c||c|c|}
\tableline
	$u_{min}$ &q &Point Source &$\rho=0.003$\\
\tableline
\tableline
0.1 & 0.01   & $94.1 \%$ & 91.3   \\
    & 0.001  & 29.3      & 16.9   \\
    & 0.0001 & 1.6       & 0.4    \\
\tableline
\tableline
0.3 & 0.01   & 59.1      & 58.8   \\
    & 0.001  & 13.8      & 15.0   \\
    & 0.0001 & 4.9       &  0.5   \\
\tableline
\tableline
0.5 & 0.01   & 34.3      & 34.9   \\
    & 0.001  & 11.3      & 8.9    \\
    & 0.0001 & 0.7       & 0.4    \\
\tableline
\end{tabular}
\end{center}
\tablenum{2}
{\bf Table 2} Lensing Zone Detection Efficiencies $\elz(q)$, $\dchit=225$: Finite
Source 
\label{tbl-2}

\bigskip

To make a quantitative comparison, we have calculated $\epsilon$ in the
same manner and for the same parameters as in \S\ 5.1, but now compare the
simulated light curves to the best-fit finite source binary light curves
with $\rho_*=0.03$.  The results with a detection criterion 
$\dchit=225$ are shown in \fig{fig:fseps}, 
along with the corresponding point source efficiencies from \fig{fig:pseps}.
In agreement with the estimate from \eq{\ref{fslimit}}, 
the detection efficiency $\epsilon$ for $q=10^{-2}$ companions 
is hardly affected.  For $q=10^{-3}$ and separations $b<1$, 
$\epsilon$ is similar to or smaller than finite source 
efficiencies, but for wider separations $b>1$, $\epsilon$ 
can be either somewhat smaller or somewhat {\it larger} 
due to the finite source size.  
The difference is dramatic for $q=10^{-4}$: 
source sizes corresponding to bulge giants always yield efficiencies 
$\lsim 1\%$.  As in \S\ 5.1, we calculate lensing zone efficiencies 
$\elz$ (\eq{\ref{elenszone}}); the results are shown in
Table~\ref{tbl-2}.  We conclude that finite source sizes have negligible 
effect on $\epsilon$ for $q\gsim 10^{-2}$, but sources as 
large as bulge giants ($\rho_*\simeq 0.03$) can have a dramatic
effect for smaller companions, 
either increasing or decreasing $\epsilon$ ($q \simeq 10^{-3}$), 
or wiping out the detection efficiency completely $(q\lsim 10^{-4})$.

Unfortunately, for individual events, the value of $\rho_*$ is very 
poorly constrained.  While it is possible to estimate the physical size
of the source from its color and magnitude, this cannot be translated to
the dimensionless projected size $\rho_*$ if the value of
$\theta_E$ remains unknown.  The detection efficiency $\epsilon$ 
for most events could be in error therefore by many tens of
percent (c.f. \fig{fig:fseps} and Table~\ref{tbl-2}).  
We discuss method of dealing with this difficulty in \S\ 8.2.

\begin{figure*}[t]
\epsscale{1.2}
\plotone{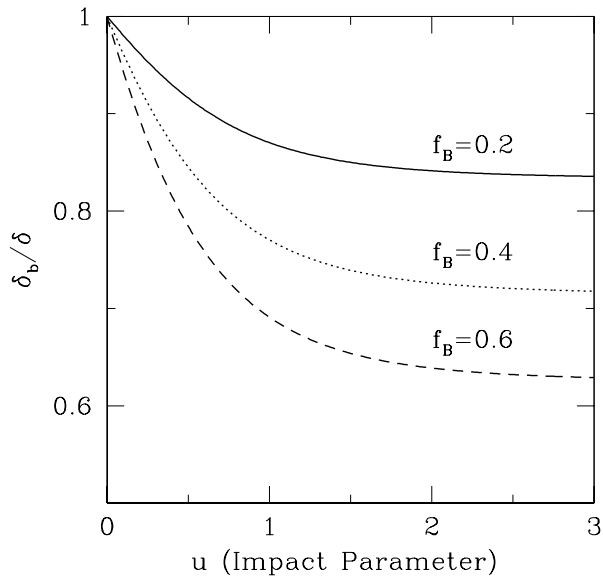}
\caption{
\footnotesize
Ratio of blended $\delta_B$ to true fractional deviation $\delta$, as a function of
impact parameter $u$ for three different blend fractions, $f_B$.  
Microlensing events are generally alerted when $u < 1$.}
\label{fig:dbrat}
\end{figure*}

\begin{figure*}[t]
\epsscale{1.5}
\plotone{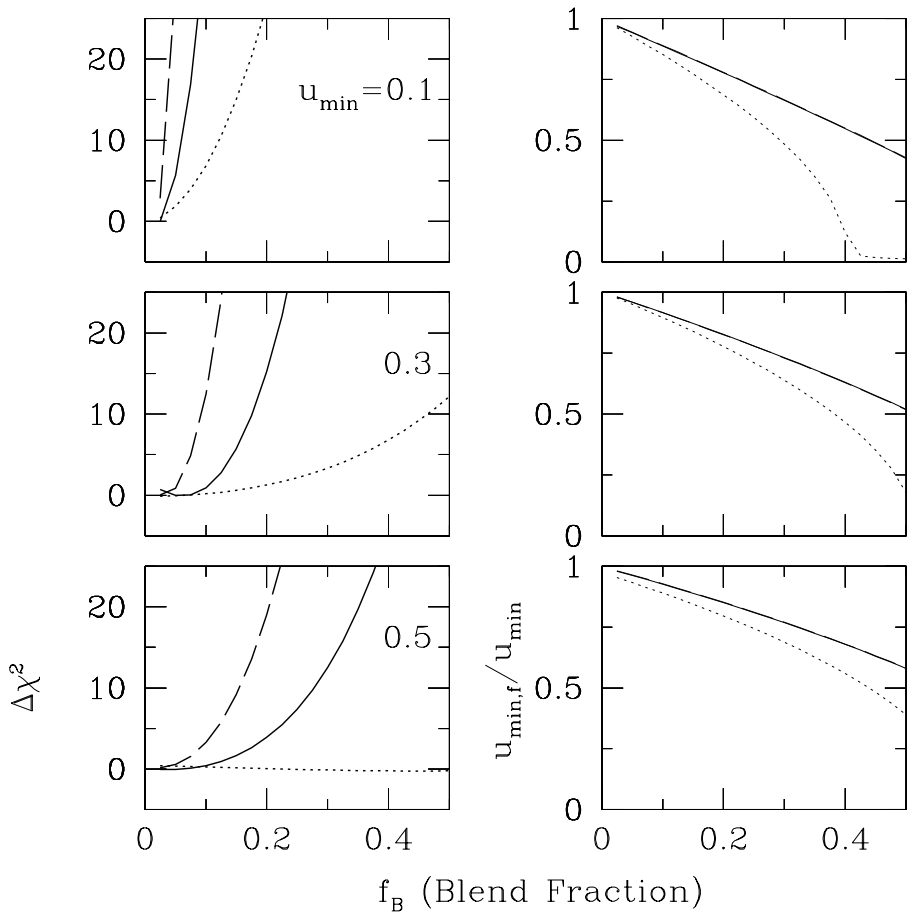}
\caption{ 
\footnotesize
Left panels: The difference in $\chi^2$ between a point-lens 
fit with no blending and one with the blend fraction held fixed 
at $f_B$, for observations from alert ($A=1.54$) until $t=\te$ 
(dotted line) and from alert until $t=3\te$ (solid line).  
The effect of improving the photometric precision by a factor 
two while monitoring to $t=3\te$ is also shown (long-dashed line).  
Right panels:  The ratio
of the fitted value of $u_{min}$ to the true value of $u_{min}$ as a 
function of the blend fraction.  Line types are the same as the left
panels; the long-dashed line is coincident with the solid line.}
\label{fig:dchi2blend}
\end{figure*}

\begin{figure*}[t]
\epsscale{1.2}
\plotone{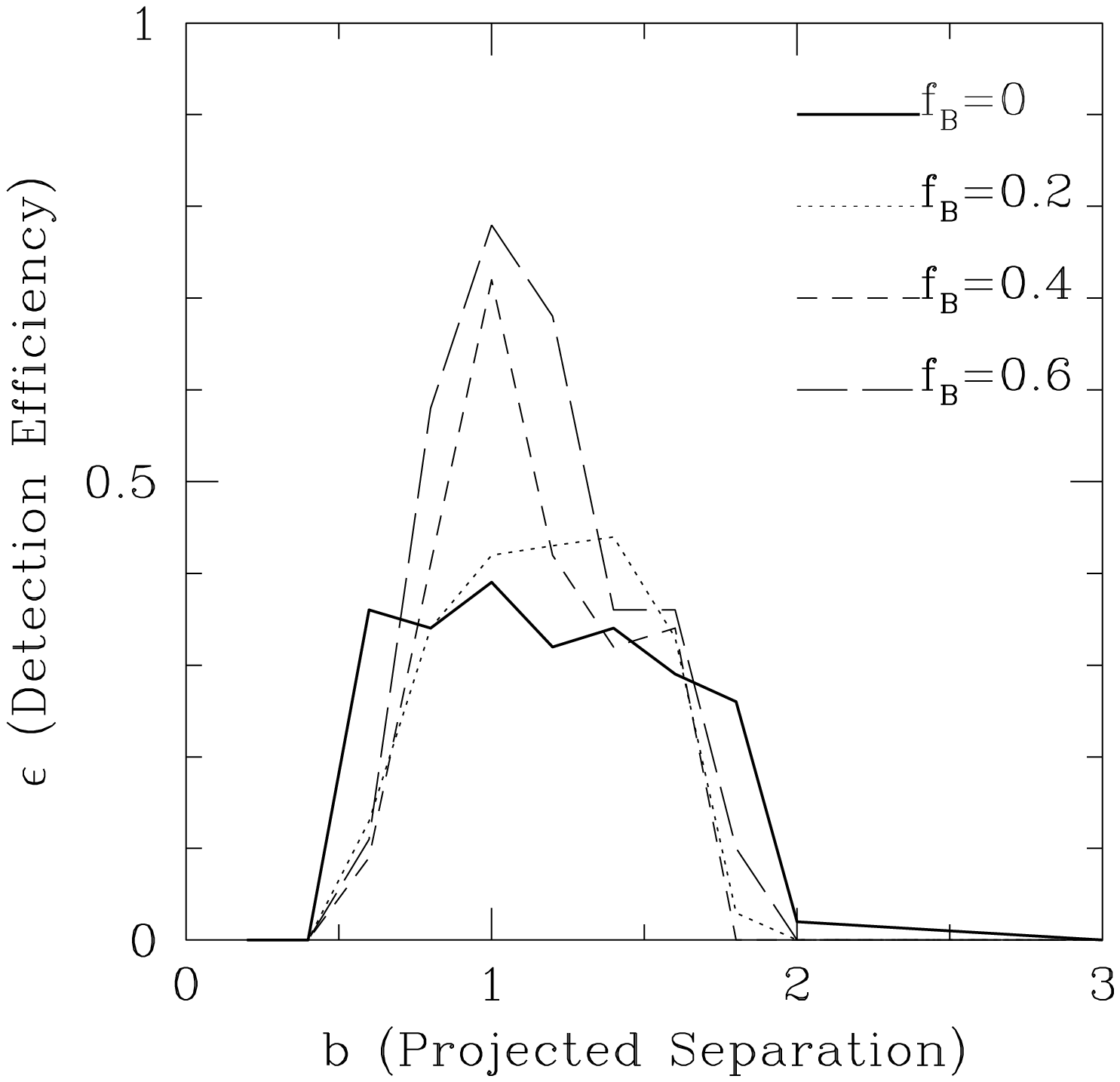}
\caption{
\footnotesize
The detection efficiency $\epsilon$ as a function of the
dimensionless separation of the binary, $b$, for mass ratio $q=10^{-2}$
and four different blend fractions $f_B$ for an event with $2\%$
photometry, uniform sampling from alert until $\te$ at the rate
$\te/200$, and minimum impact parameter $\umin=0.5$.}
\label{fig:bleps}
\end{figure*}

\section{Blending}

A microlensing event is blended whenever unresolved, unlensed background
light contributes significantly to the 
observed baseline flux of the source star (i.e. $f_B \ne 0$).  
From \eq{\ref{fluxt}}, the observed
magnification in the presence of blending, $A_B$, is related to the true
magnification by,
\begin{equation}
A_B= (1+f_B)^{-1}(A + f_B)~,
\label{blendmag}
\end{equation}
where $f_B$ is the ratio of the blend flux to the true lensed source flux.

Blending will have two effects on the detection efficiency.  The first
is a suppression of the deviation caused by the binary.  From \eq{\ref{blendmag}} and \eq{\ref{fluxt}}, it is straightforward to
show that the fractional deviation in the presence of a blend,
$\delta_B$, is related to the true fractional deviation by,
\begin{equation}
\delta_B =\delta \left({ 1 + {f_B \over A_0}}\right)^{-1}~.
\label{fracblend}
\end{equation}
As before, $A_0$ is the magnification of the best-fit PSPL light curve.  
Note that
\eq{\ref{fracblend}} applies to any anomaly that produces a deviation from the standard PSPL light curve,
including parallax, binary source, and finite source effects.  Since
$A_0$ is a function of time, the magnitude of the suppression will also
be a function of time, such that deviations occurring closer to the peak
($u \sim \umin$) will be less suppressed than those occurring near the
beginning and end of the event.  \fig{fig:dbrat} shows the ratio $\delta_B /
\delta$ as a function of the angular separation $u$ of the lens and
source, for three values of $f_B$ corresponding to relatively mild
blending, $f_B < 1$.  Anomalies occurring near the peak of high
magnification events ($u\sim 0 $) will be only slightly suppressed 
(\cite{gands1998}), while repeating events caused by wide binaries 
(Di Stefano \& Mao 1996; Di Stefano \& Scalzo 1999b) can be suppressed
by as much as $(1+f_B)^{-1}$.  Overall, the amplitude of the suppression
for mild blending is relatively small, $\delta_B/\delta \le (1+f_B)^{-1}$,
and thus will not have a large effect on the detection efficiencies 
for binaries with mass ratios $q \ge 10^{-2}$ 
for which the fractional deviations are usually large.  
For mass ratios consistent with planets, $q \le 10^{-3}$, however, a
substantial fraction of detected events will have maximum fractional
deviation $\delta_{\rm max} \lsim 0.1$ so that even a suppression of $\sim
0.5$ can have a significant effect on planetary detection efficiencies.

The second effect that blending has on detection efficiencies is to
alter the presumed distribution of $\umin$ in observed events.  The
intrinsic distribution of $\umin$ is flat up to the magnification
threshold set for detection by the survey teams (e.g., $A>1.34$
corresponds to $\umin<1$).  This threshold is calculated in real time
near the beginning of the event when a robust determination of blending
is not possible.  Consequently, the magnification of the event at any
time is assumed to be the total flux divided by the baseline flux,
$A_B=(1+f_B)^{-1}(A+f_B)$, which is less than the true magnification $A$ for
all non-zero blending values.  Thus, in the presence of blending, an
event will require a larger magnification and thus smaller intrinsic
$\umin$ in order to pass the detection criterion, which is {\it per
force} applied to the observed quantity $A_B$ (see \cite{alcock1997a}
for an example and discussion).  Since the intrinsic detection
efficiency is larger for smaller impact parameters (see \S\ 5.1 and
\fig{fig:pseps}), the blended event will be more sensitive to the presence of
planets than would be calculated for a $\umin$ based on the (erroneous)
assumption that $f_B=0$.  Blending thus affects binary detection efficiency
in two competing ways: the efficiency is decreased by suppressing 
the amplitude of observed deviations, while at the same time it is 
increased due to the skewing of the observed $\umin$ distribution 
to smaller values.
The net effect on the detection efficiencies will depend on the values of
$b$ and $q$ and vary on an event-by-event basis.

Clearly, blending must be considered when calculating binary and
planetary detection efficiencies.  
Since blending effects are relatively easy to quantify, 
this poses no serious complication as long as the blending parameter, 
$f_B$, can be accurately determined for individual events.   
Unfortunately, as discussed by Wo{\'z}niak \&
Paczynksi (1997), blending can be extremely difficult to determine for
individual observed light curves, due to the serious correlations in the
parameters $F_0, \te, \umin$ and $f_B$ in the presence of blending. The
degeneracy is especially severe in two regimes.  In the `spike' regime,
defined by $f_B \gg 1$ and $\umin \ll 1$, 
$\te$ and $\umin$ cannot be measured separately and only the
degenerate combination $\te/\umin$ is measurable (\cite{gould1996}).  
This regime is most
important in severely crowded fields, such as those towards M31.  
The second regime, defined by large impact parameter 
($\umin \gsim 0.5$) events with modest blending $f_B < 1$, 
is more common in bulge fields, and thus the focus of our current attention. 

In order to illustrate the difficulty in quantifying the fraction of
blended light, we calculate the range of allowed values inferred 
for $f_B$ from fits to our fiducial simulated light curves 
(i.e., observations from alert until $\te$, $2\%$ errors, and 
{\it no blending}).  Fixing $f_B$ at some value,
we find the best fit to a point-lens model allowing the parameters
$\umin, \te, t_0,$ and $F_0$ to vary.  We then compute $\dchi$ between
this fit and that assuming $f_B=0$.  
The resulting $\dchi$ as a function of $f_B$ is shown in
\fig{fig:dchi2blend}.  For $\umin=0.1$ and $\umin=0.3$, the degeneracy is not
severe; the $3\sigma$ allowed ranges in $f_B$ are $f_B \le 0.125 $ and
$f_B \le 0.45$, respectively.  
For $\umin=0.5$, the blend fraction is almost completely unconstrained
because for large $\umin$ events constraints on blending arise mostly
from the combination of information from the wings 
($|t-t_0| \sim \te$) of the event and the
baseline ($|t-t_0| \gg \te$).  Thus, without a baseline measurement, the value of $F_0$ can
be arbitrarily adjusted to compensate for large values of $f_B$ without
significantly affecting the fit.  With this in mind, we have also
computed the same $\dchi$ statistic for a simulated light curve with
observations from alert until $3 \te$.  Here the blending is much better
constrained, $f_B \le 0.27$, with additional improvement if
the errors are reduced by half, in which case $f_B \le 0.15$.  
Since the majority of the constraint comes from sampling the wings and
baseline of the light curve, it would be more efficient for 
monitoring teams to concentrate on more, rather than better, measurements.

The right panels of \fig{fig:dchi2blend} demonstrate how an 
inaccurately-determined blend fraction can affect the 
determination of $\umin$ and thus the detection efficiency 
$\epsilon$.  Here we show the ratio between the value of $\umin$
determined by assuming a constant blend fraction $f_B$, $u_{\rm min,f}$,
and the true value $\umin$.  The value of $\umin$ deduced for 
blend fractions between $f_B=0$ to $f_B=0.4$ can vary by nearly 
$50\%$ resulting in quite different inferred detection efficiencies 
(c.f. \fig{fig:pseps}).  
To quantify this, we have calculated $\epsilon$ as a function of $f_B$ for
an extreme example, $\umin=0.5$ and observations from alert until
$1\te$.  The procedure for calculating $\epsilon$ is the same, except
now $f_B$ is fixed at an assumed value and $F_0$ is included as a free
parameter in both the single lens and binary lens fits.  The results are
shown in \fig{fig:bleps}, where we plot $\epsilon(q=10^{-2})$ as a function of
$b$ for $f_B=0$ (same as \fig{fig:pseps}), $0.2$, $0.4$, and $0.6$.  
Recall that all these blend fractions are statistically indistinguishable 
for this light curve. The differences in $\epsilon$ are dramatic.  The 
suppression of binary anomalies induced by blending of $f_B=0$ and $0.2$ 
causes a drop in detection efficiency for separations 
$b \lsim 0.6$ and $b\gsim 1.6$.
Inside the Lensing Zone, however, the net effect is a dramatic increase in
$\epsilon$ due to the lower value of $\umin$ required to produce the
observed light curve for increasing $f_B$.  This can be appreciated 
best by examination of Table~\ref{tbl-3}, which tabulates $\elz$ 
as a function of $f_B$.  To the extent that
these values of $f_B$ cannot be distinguished from one another by the
light curve alone, they are all equally likely, and thus $\epsilon$ can
be quite uncertain.

\bigskip

\begin{center}
\begin{tabular}{|c||c||c|}
\tableline
	$f_B$ & $u_{min,f}$ & $\elz, \dchit=225$ \\
\tableline
\tableline
        0.0 & 0.50      &  34.3\%  \\ 
        0.1 & 0.44      &  32.7    \\
        0.2 & 0.40      &  37.2    \\
        0.3 & 0.34      &  39.6    \\
        0.4 & 0.28      &  41.7    \\
        0.5 & 0.19      &  56.8    \\
        0.6 & 0.02      &  52.7    \\
\tableline
\end{tabular}
\end{center}
\tablenum{3}
{\bf Table 3} Lensing Zone Detection Efficiencies $\elz(q=10^{-2})$:
Blending 
\label{tbl-3}

\bigskip

Blending can give rise to a serious uncertainty in derived detection
efficiencies but, unlike finite source size, the blend fraction can be
determined with sufficient accuracy for most events.  To do so, however,
requires precise measurements by the monitoring teams 
during the wings of the event and at baseline.  Without a reasonable
quantification of the blend fraction, the detection efficiency of 
individual events will be very uncertain.

\section{Application to Real Data}

In previous sections, we used artificial data to explore several effects 
that can influence significantly the 
determination of the detection efficiency $\epsilon$ of individual 
light curves, including detection criteria (\S\ 5.1), 
finite source size (\S\ 6), and blending (\S\ 7). 
These effects are often difficult to quantify in real data, for
which sampling and photometric precision is likely to vary on an
event-by-event basis, and with observing conditions and microlensing
phase for individual events.  Furthermore, real-time observational
decisions may alter (increase) the sampling of clearly anomalous events
from that of apparently non-anomalous PSPL events.

\subsection{Variable Sampling and Photometric Precision}

Since the algorithm we presented in \S\ 5 uses the actual light curve, 
and thus the actual sampling and photometric uncertainties associated with 
each observed event, irregular sampling and variable precision are 
taken into account explicitly in the determination of the 
detection efficiency $\epsilon$.  
The efficiencies based on artificial data presented in previous sections 
to illustrate general principles will not be strictly applicable 
to real microlensing events, for which weather and other considerations 
prevent continuous monitoring with a sampling of $\te/200$.  
In general, the effect of reduced or incomplete sampling will be
to lower detection efficiencies.  For extremely long events, or those
(including very short) events alerted post-peak by the survey teams,
substantial portions of the light curve will have no (dense) monitoring
at all, and the effects on the detection efficiency can be quite
devastating.  In such partially-monitored light curves, the PSPL fit
parameters ($\umin, \te, t_0, F_0$ and $f_B$) will be very uncertain, and 
in extreme cases almost completely unconstrained.  Since the detection
efficiency depends on these parameters, the resulting uncertainty in
$\epsilon$ will be quite large.  Unless additional information is
available (e.g. from the survey teams) to constrain the fit, these data
will add almost nothing to our knowledge of the abundance of planets
since their detection efficiency cannot be reliably quantified.

For Gaussian, uncorrelated errors, the $\chi^2$ statistic can be used 
as a measure of goodness-of-fit.  Real
measurement uncertainties are seldom truly Gaussian, especially in 
crowded microlensing fields where systematic effects associated with seeing,
scattered light, and detector characteristics become increasingly
important.  Uncertainties on individual data points are often taken to be the
formal errors reported by the PSF-fitting algorithms of photometric
reduction packages like DoPhot (\cite{dophot}), which often underestimate
the true scatter (\cite{albrow1998}).  Image subtraction techniques
(\cite{tandc1996}; \cite{andlup1998}) may alleviate some of
these difficulties, but for the moment are too cumbersome and slow to
implement for multi-site, real-time reduction of large fields and thus
have not yet been implemented by monitoring teams.  An empirical
correction to account for the correlation of measured photometric
magnitude with the FWHM of the point spread function often results in a
more Gaussian error distribution whose average magnitude corresponds
more closely to the formal DoPhot-reported error 
(Naber, private communication; \cite{planetOB9814}).  
As long as the detection criterion $\dchit$ is
maintained at a suitably high value (\S\ 5), the exact error
distribution is likely to have little effect on the computed efficiency
$\epsilon$.  Remaining doubts can be assuaged by attaching the observed
error distribution derived from constant stars to artificially generated
PSPL light curves to calibrate both the `false alarm rate' and 
efficiency of true detections with a given $\dchit$ criterion.

\subsection{Finite Source Effects}

Finite source effects pose a significant challenge to the robust
determination of the detection efficiency $\epsilon$ because the
dimensionless source size, $\rho_*$ is unknown a priori.  As we have
shown in \S\ 5, the finite size of the source
should have negligible effect on $\epsilon$ for $q \gsim 10^{-2}$
but can have a significant effect for $q \lsim 10^{-3}$.
Thus, without additional information about $\rho_*$, $\epsilon$ can be
determined robustly only for $q \gsim 10^{-2}$, which is unsatisfactory 
for microlensing monitoring programs whose primary goal is to learn about 
small mass-ratio systems.  A first-order estimate for $\rho_*$ 
could be obtained by measuring the angular size of the source, $\theta_*$, 
and assuming that the relative proper motion of the lens, 
$\mu = v_\perp / \dol$, is equal to the mean relative proper motion
$\langle\mu\rangle$ for all lenses toward the bulge.  
The dimensionless size of the source would then be given by 
\begin{equation}
\rho_* \approx {\theta_* \over \langle \mu \rangle \, \te}~.
\label{rhoprob}
\end{equation}
The angular size of the source can be
estimated by its color and magnitude, or by obtaining a precise 
spectral type through more resource-intensive
spectroscopy. Unfortunately, the value of $\langle\mu\rangle$ depends on the
assumed velocity and spatial distribution of the lenses.  Moreover, 
even within a given model, the distribution of $\mu$ is wide, having a
variance of a factor of $\sim 3$ and long tails toward higher and
lower values (\cite{handg1995}).  The true value of $\mu$ thus 
could differ substantially from $\langle\mu\rangle$, leading to a large error
in $\epsilon$ for individual events.  When averaged over 
many events, these errors should approximately cancel, but 
current monitoring programs are very far from this regime, 
especially for {\it anomalous\/} events.

A somewhat better estimate for the effect of $\rho_*$ on 
detection efficiencies could be made as follows.  
Assume that an event
has measured time scale $\te$ and angular source size $\theta_*$. 
For an assumed model of lens distances and velocities, the expected 
distribution of proper motions, ${\cal G}_i (\mu)$, can be computed. 
The individual detection efficiency
can then be approximated as,
\begin{equation}
\epsilon_i(b,q)= \int_{\mu_{\rm min}}^{\mu_{\rm max}} {\rm d}\mu\,
\epsilon_i(b,q;\rho_*) {\cal G}_i (\mu) ~,
\label{efsource}
\end{equation}
where $\rho_*={\theta_* / \mu\te}$, $\mu_{\rm max}$ is the 
maximum proper motion allowed by the
observed light curve, and $\mu_{\rm min}$ is some reasonable lower limit.  
This model-dependent estimate of the detection efficiency should 
be more accurate whenever finite source effects are large, but it 
is time consuming to compute because $\epsilon(b,q)$ must be determined for
many $\rho_*$.

Ideally, one would like to determine $\rho_*$ directly for each 
individual lens.  This can be done by measuring $\mu$ from 
single-color light curves for only $\sim 5\%$ of events
(\cite{gould1994}, \cite{witt1995}), 
namely those with high peak magnification.  
With both optical
and infrared photometry, $\rho_*$ could be determined for approximately
twice as many events (\cite{gandw1996}; \cite{witt1995}). 
If the lens is luminous,
one could measure the proper motion of the lens directly using accurate
astrometry and a high resolution instrument, such as HST
(\cite{gandg1997}).  
This will not be possible for all events, however, and requires
a long temporal baseline.  The angular Einstein ring radius 
$\theta_{\rm E}$ can be determined directly by measuring the 
centroid shift of the two unresolved images created by the lens.  
As the lens passes across the
line of sight to the source, these two images move and change
magnification.  The centroid of these two images traces out an
ellipse who size is $\sim \theta_E$ (\cite{walker1995}), thus requiring an   
an astrometric accuracy considerably smaller than 1~mas.  
Preliminary studies have shown that SIM, with its
planned $4\mu {\rm as}$ accuracy, should be able to measure
$\theta_E$ reliably for almost all known microlenses toward the galactic bulge
(\cite{boden1998a}; \cite{jeong1999}; \cite{dandsahu1999};
\cite{samir99}).  
Ground-based interferometers
currently being developed, such as the Keck Testbed Interferometer,
should be able to measure $\theta_E$ for a smaller, but substantial,
fraction of events (\cite{boden1998a}).  Measurements of this kind would
require coordination and cooperation between microlensing and
astrometric communities, but the results would be well worth the effort.

In summary, if microlensing monitoring teams can make use of other
resources, such as HST and SIM, the best method of dealing with finite
source size effects is simply to measure the dimensionless source 
size $\rho_*$ for each individual event directly.  In the
absence of these options, the effect of finite source size on $\epsilon$ 
must rely on statistical and model-dependent estimates of the distribution 
of $\rho_*$.

\subsection{Blending}

As discussed in \S\ 7, the accurate determination of the blending, $f_B$,
in individual light curves is essential to the accurate determination of
the detection efficiency.  Precise measurements during the
wings of the event and at baseline can allow the quantification of
$f_B$, but may not always be possible: bad weather may prevent  
wing measurements in some events and the faintness of the source star 
may preclude precise baseline measurements in others.  
Improved or alternate methods of quantifying blending would be beneficial; 
several have already been suggested.

Since the blending problem arises mainly because the PSFs of
individual stars are blended together in crowded microlensing fields, 
simply improving the spatial resolution of the observations
eliminates or reduces the blended light in most cases.  
The resolution needed is roughly that of HST (\cite{han1997}), for which 
only a modest use of resources is required.  
Nevertheless, even the resolution of HST will be insufficient 
to resolve any blended light that might arise from the lens itself or 
unresolved companions to the lens or source.  
Blending caused either by the lens or 
unrelated projected stars near the source or lens 
will generally cause a color shift in the combined light as the 
source is magnified during the course of the event (\cite{bandk1996}).  
Since most stars in the bulge have nearly the same color, the color
shift is expected to be small and difficult to measure.  
Alternatively, the blend fraction can be quantified by measuring the 
centroid shift of the blended PSF during the course of the event\footnote{
This is not to be confused with the ${\cal O} (1 \, {\rm mas})$ centroid shift 
caused by the 
motion and variable magnification of the two images created during 
a microlensing event.
The centroid shift due to blending is $\sim d (A_B-1)/A_B$, where $d$ is the
separation of the blended sources and $A_B$ is the observed (blended)
magnification.  Since $d$ is of order the
resolution, $1"$, this centroid shift is measurable using current
ground-based observations.} (\cite{gandw1998}; \cite{gberg1998}).
Blend fractions can be reliably determined in this way only for 
heavily blended, high-magnification events (\cite{hjk1998}) in which 
the centroid shift is prominent.  
Finally, improvement can be made through using image subtraction
rather than the usual PSF-fitting photometry.  Image subtraction 
reveals only the time-variable flux, $\Delta F=F_0[A(t)-1]$, so that constant sources of 
blended light are removed.  Although image subtraction does not remove
the correlations between fitted parameters that make blending
problematic, it is useful since photometric uncertainties are
generally near the photon noise 
limit (\cite{tandc1996}; \cite{andlup1998}), which allows better 
discrimination between subtle differences in blended light curves.

\subsection{Anomalous Events}

If solar systems like our own are not atypical, dense, precise 
photometric monitoring will eventually result in the detection of 
anomalous events consistent with planetary perturbations.  The 
algorithm of \S\ 5 can be used to place direct limits on planetary
systems in non-anomalous events and give meaning to the large
number of null results now in hand, but what of the detection efficiency
of events in which planetary anomalies are actually observed?

By the detection efficiency $\epsilon_i(b,q)$ of an anomalous event, we
mean, by analogy with the definition for non-anomalous events given in
\S\ 4, the probability that a lensing companion with parameters $b,q$
would be detected in light curve $i$ 
(using the {\it same} $\dchit$ criterion) after
integrating over source trajectory angles $\theta$.  In
principle, if the parameters $t_0, \te, \umin, F_0$ and $f_B$ can be
well-determined in a binary fit that meets the $\dchit$ criterion,
$\epsilon(b,q)$ can be calculated in a manner similar to that presented
in \S\ 5.  A PSPL light curve with these parameters can be generated
with the {\it actual sampling and photometric errors} of the observed
event and the method of \S\ 5 applied for the $b,q$ combination that
produced the best-fit `detected' binary.

Several difficulties inherent to anomalous light curves must be 
addressed.  First, for large mass
ratio ($q\gsim 10^{-2}$), the source trajectory $\theta$ may actually be
constrained by the light curve itself, but in a way that is partially
degenerate with other fitting parameters.  Second, even for small mass
ratios, i.e., true planetary anomalies, the parameters $t_0, \te, \umin,
F_0$ and $f_B$ may not be well-constrained, though of course this
difficulty is present for non-anomalous events as well.  Third, the
anomaly, if large, may produce changes in magnification significant
enough to alter the photometric precision obtained at the phase of the
perturbation.  Most planetary anomalies will have rather gentle changes
in magnification (\fig{fig:lcurves}) so that this effect may not be severe for
$q\lsim 10^{-3}$.  Fourth, if detected real time, observer intervention
may alter the photometric sampling in a way that is also
phase-dependent (i.e. higher sampling after the anomaly detection than
before).  These last two possibilities present difficulties when integrating
the corresponding PSPL light curve over source trajectory $\theta$,
since the altered sampling and photometric errors will not correspond to
the phase of actual anomalies for most choices of $\theta$.  

One approach to handling this increased sampling at the phase of the
anomaly is to resample the `post-anomaly' portion of the light curve in a
manner that is consistent with what would have been the sampling had the
anomaly gone unnoticed.  (Determining how to do this in the presence of
weather and other observing facts-of-life may not be trivial).  Both the
initial test of binarity and the procedure to determine the detection
efficiency would then be performed with the sparser, resampled light
curve.  The full dataset would be used only to refine the best-fit
parameters.  Alternatively, a Bayesian approach to analyzing the full
dataset (anomalous and non-anomalous) could be employed, which addresses 
many of the special difficulties posed by anomalous events
(\cite{sackett1999b}).

\section{Conclusion}

In this paper, we have presented and tested a method to calculate the
detection efficiency of microlensing datasets to stellar and planetary lensing 
companions.
This method is conceptually simple, direct, and not excessively time
consuming.  The final result, $\epsilon(b,q)$, conveniently summarizes
the efficiency of a given dataset in revealing binary lensing 
systems with any mass ratio $q$ and dimensionless separation $b$.  
This efficiency can then be used to evaluate the
likelihood that any given model of planetary systems would give rise to
the observed dataset.

The algorithm that we have presented has several advantages.  
First, by working directly with measured light curves, it makes 
few assumptions, and explicitly takes into account the dependence 
of the efficiency on the sampling,
photometric precision, monitoring duration, and impact parameter of
the events.  Second, it is computationally efficient, involving 
direct integration over only one unknown parameter, the angle of the
trajectory.  Finally, it is convenient, incorporating all 
information into one two-dimensional function, $\epsilon(b,q)$, which
can then be used to evaluate the likelihood of any model of planetary
systems.

Some caveats must be noted.  First, the method
assumes that the primary lensing parameters $t_0, \te, \umin, F_0$ and
$f_B$ are well-constrained by the light curve, while emphasizing that
$\epsilon$ is poorly known in the absence of such constraints.  
Possible altered sampling and photometric precision of anomalous events 
may necessitate a resampling of the corresponding light curve or the 
assumption that the efficiency of anomalous events
is equal to the average efficiency of the normal
events.  Resampling is not an efficient use of all the data, and for
individual events the average efficiency of non-anomalous events is
likely to be a rather poor proxy for $\epsilon_i$.  
Finally, the algorithm implicitly assumes that the efficiency is 
independent of the planetary system model 
(e.g., the effect of multiple planets is ignored).

We have also applied our method to artificial light curves in 
order to explore the dependence of binary detection efficiency on the 
impact parameter of the event, the choice of
the detection criterion, neglecting to fit comparison binary curves 
separately, finite source effects, and blending.  Our conclusions are as
follows:
\begin{description}
\item[{(1)}] Most of the efficiency of microlensing datasets 
to detecting planetary anomalies results from non-caustic crossing events.
\item[{(2)}] Detection efficiency is strongly dependent on the impact parameter
$\umin$ of the event.  This implies that the integrated detection
efficiency will depend strongly on the actual distribution of impact
parameters of the monitored events.
\item[{(3)}] For mass ratios $q\ge 10^{-2}$, the inferred detection 
efficiency $\epsilon$ is robust to changes in the detection 
criterion $\dchit$. For smaller mass ratios $q\sim 10^{-4}$ 
near the detection limit, different choices of $\dchit$ can lead to 
differences in $\epsilon$ of a factor of two. 
\item[{(4)}] In order to obtain accurate detection efficiencies, 
the binary lens light curve must be fit separately unless the mass ratio 
is small ($q\lsim 10^{-3}$) and the deviations are well above 
the detection limit (which may be difficult to satisfy simultaneously).
\item[{(5)}] Finite source effects are negligible for $q \gsim 10^{-2}$, can either
increase or decrease $\epsilon$ for $q \simeq 10^{-3}$, and can be
devastating for $q \lsim 10^{-4}$, at least for the large dimensionless
source sizes ($\rho \gsim 0.03$) that are typical of bulge giants.
\item[{(6)}] The detection efficiency is very sensitive to the fraction of blended light,
primarily due to item (2) above; higher blend fractions imply smaller
impact parameters for detection, and smaller impact parameters have
higher intrinsic detection efficiencies.  This fact, combined with the
degeneracy in the blend fraction that exists between poorly/inaccurately
sampled blended light curves means that {\it monitoring teams should
make every effort to quantify the blend fractions in every monitored
light curve}.
\end{description}

The final result of applying this method to an observed data set is the
integrated detection efficiency, $\epsilon(b,q)$.  This result should be
regarded as a primary outcome of microlensing monitoring teams
conducting planet searches; it incorporates all the information from the
observations, and is essential for establishing any conclusions about
the abundance and nature of Galactic planetary systems.

\acknowledgements

It is a pleasure to thank Darren Depoy, Martin Dominik, Andy Gould,
Konrad Kuijken, Richard Naber, David Weinberg, and Simon Wotherspoon for
stimulating discussions and comments on the manuscript.  
One of us (B.S.G.) would like to acknowledge
the support and generosity of everyone at the Kapteyn Institute, where
the majority of this work was completed.  
This work was supported in part by NASA under Award No. NAG5-7589, 
the NSF under grants AST~97-27520 and AST~95-30619 and by a visitor's grant
from Kapteyn Institute.

\end{document}